\documentclass[5p,times,number]{elsarticle}
\usepackage[english]{babel}
\usepackage{graphicx}
\usepackage[breaklinks, colorlinks, citecolor=blue]{hyperref}
\usepackage{subfigure}
\usepackage{amsmath,amssymb}
\usepackage{xspace}
\usepackage{cleveref}
\usepackage{mathrsfs}
\usepackage{booktabs}

\newcommand{\refeq}[1]{\cref{eq:#1}}
\newcommand{\sect}[1]{sect.~\ref{sec:#1}}
\newcommand{\Fig}[1]{Fig.~\ref{fig:#1}\xspace}
\newcommand{\lgg}[2]{\ensuremath{\mathscr{L}_{#1}(#2)}}

\newcommand{\mean}[1]{\ensuremath{\langle#1\rangle}}
\providecommand{\norm}[1]{\lVert#1\rVert}

\newcommand{\nn}{\nonumber}
\providecommand{\E}[1]{\ensuremath{\mathbb{E}\left[#1\right]}}

\newcommand{\hosp}{\textsl{hosp}\xspace}
\newcommand{\conf}{\textsl{conf}\xspace}
\newcommand{\malawi}{\textsl{malawi}\xspace}
\newcommand{\baboons}{\textsl{baboons}\xspace}
\newcommand{\Nclus}{\ensuremath{N_{\text{c}}}\xspace}
\newcommand{\bern}[1]{\ensuremath{\mathscr{G}(#1)}}

\newcommand{\ccdf}{CCDF\xspace}
\newcommand{\CLT}{CLT\xspace}

\newcommand{\citeg}[1]{(see e.g., \cite{#1})}

\newcommand{\SI}{\textit{SM}}
\newcommand{\Sup}[1]{\SI \textit{-#1}\xspace}
\newcommand{\safeurl}[1]{\href{#1}{#1}}

\begin{document}

\begin{frontmatter}

\title{A stochastic model of discussion}

\author{S. Plaszczynski \corref{cor1}}
\ead{stephane.plaszczynski@ijclab.in2p3.fr}
\author{B. Grammaticos} 
\author{M. Badoual}

\address{Universit\'e Paris-Saclay, CNRS/IN2P3, IJCLab, 91405 Orsay, France}
\address{Universit\'e Paris-Cit\'e, IJCLab, 91405 Orsay France}
 
\cortext[cor1]{Corresponding author}
\begin{abstract}
We consider the duration of discussions in face-to-face contacts and
propose a stochastic model to describe it. It is based on the points of
a Levy flight where the duration of each contact corresponds to the size
of the clusters produced during the walk. When confronting it to the data
measured from proximity sensors, we show that several datasets
obtained in different environments, are precisely reproduced by the
model fixing a single parameter, the Levy index, to 1.15. 
We analyze the dynamics of the cluster formation during the walk
and compute analytically the cluster size distribution. We find that discussions
are first driven by a maximum-entropy geometric distribution and then
by a rich-get-richer mechanism reminiscent of preferential-attachment
(the more a discussion lasts, the more it is likely to continue).
In this model, conversations may be viewed as an aggregation process
with a characteristic scale fixed by the mean interaction time between
the two individuals.
\end{abstract}


\end{frontmatter}

\section{Introduction}
\label{sec:intro}

Communication networks as face-to-face ones are described by 
time-varying graphs which are difficult
to characterize because of the intertwining of the topological and
temporal aspects  \cite{holme:2012,holme:2015}. 
However, the sociological mechanisms that drive people encounters
may actually be different from the ones that are involved
when two individuals actually ``interact''. Since in the following we only
consider cases where people are face-to-face for at least 20 s, this
means for humans
that a \textit{discussion} is engaged. It could follow some
universal rules that lie out of the social context 
as stipulated in the field of Conversation Analysis \cite{Button:2022}.
This is suggested by a recent analysis
\cite{Plaszczynski:2024}
of the \textit{sociopatterns} data (see \Sup{S1} for their
description).
It was shown there that while the mean-time we spend together
with a given person varies according to our preferences but also to the
sociological context, the \textit{deviations} from it seem to present
a universal distribution. Most often we tend to allocate short times to our
exchanges with a given individual (with respect to the usual
mean-time) but very long discussions may emerge.
What is remarkable is that the statistical distribution of this
deviation, in particular for large values, 
looks the same in very different environments as 
between scientists at a conference or between farmers in a small
Malawi village. This could point to some general properties of human face-to-face
interactions, or more generally of primates since similar results
were observed in a baboon population \cite{Plaszczynski:2024}.

We try to model here the invariant distribution observed on what was
called the ``contact duration contrast'' (or simply ``contrast'')
which is simply, for each relation (pair of individuals), the duration of contact divided by its mean-time.
We have little indication on how to build such a model but we may rely on
a very common mechanism that was dubbed ``preferential attachment'' in
network science \cite{Barabasi:1999}.
Introduced to explain the emergence of power-law degree
distributions in real-life networks 
\cite{Barabasi:1999b,Krapivsky:2000,Dorogovtsev:2000}, 
 it has in fact been studied by scientists under different forms since a
 century \cite{Yule:1925,Simon:1955, Price:1976}. 
It is even known in common language, as  ``Matthew effect''
\footnote{Because of the sentence in the Bible ``
For unto every one that hath shall be given, and he shall have
abundance [...] '' (Matthew 25:29)}. 
It formalizes the idea of the ``rich get richer'' effect. 
For instance 
\begin{itemize}
\item when a paper is abundantly cited, it is more likely to be known by
  others and thus cited even more \cite{Price:1976,Barabasi:1999}. 
  In the same vein, the more someone gets social rewards the
  more it is likely to be known by others and receive even more \cite{Merton:1968}.
\item in the so-called Yard-Sale model \cite{Hayes:2002} a pair
of agents bets a fraction of wealth of the poorest of the two
with \textit{equal probability to win}. Running such a system with
several agents, the pairs being randomly selected, finally leads
to ``oligarchy'', i.e. a single agent concentrates all the wealth
\cite{Boghosian:2015}. This is 
due to a rich-get-richer effect; when there is imbalance, it is less risky for the rich to bet some
 small amounts than for the poor, which further accentuates
 imbalance.
\item in colloidal solutions, the more an aggregate grows, the more it
  is likely to be hit by particles and become even larger.
\end{itemize}
Discussions also hold a preferential-attachment mechanism. Everyone
experienced that \textit{when a conversation lasts for long, it is likely
to last even more}. One can hardly escape it since there are 
more and more potential topics to come back to.

We propose a model to describe the dynamics of the
discussion evolution, initially motivated by a 
coincidence on the shape between two distributions.  It is based on
geometric graphs build from the points of a Levy walk
\cite{Plaszczynski:2022} that are 
first reviewed in \sect{levy}. Then we tune our empirical model
to investigate at which level it can match the \textit{sociopattern}
data in \sect{data}. This single-parameter model gives excellent
results. Investigating the dynamics of clusters formation 
we show analytically why in \sect{dynamics}.
Since the process we propose
resemble that of a growing network \cite{Krapivsky:2001}, we compare
them in more details in \sect{agg}. 
Finally in \sect{disc}, we comment on the geometric nature of the model
and emphasize why, in the class of all random-walks, only Levy flights
can be suitable candidates to model discussions through a geometric graph. 
\ref{app} explains what the self-similarity
of a Levy flight precisely means, and why it gives rise to several scaling relations in Levy graphs.
Some auxiliary results are available in the Supplementary Material (\SI).

\section{Levy flights, graphs and clusters}
\label{sec:levy}

A Levy flight, as introduced by B. Mandelbrot \cite{Mandelbrot:1975, Mandelbrot:1983}
is a random-walk process but unlike the standard case where each step
follows a Gaussian distribution (or converges to it due to the Central Limit
Theorem, \CLT), the length of the jumps follow a power-law
distribution, namely a Pareto-Levy one \cite{Mandelbrot:1960}.
The probability for a step length to exceed some value $R$, i.e. the complementary cumulative
density function (\ccdf) is 
\begin{align}
\label{eq:LF}
P(r>R)=
  \begin{cases}
    1  & \text{if}~ r \leq r_0, \\
    (r_0/R)^\alpha & \text{otherwise.} 
  \end{cases}
\end{align}
$r_0$ is a cutoff that ensures the proper probability density function
normalization, 
and $0\leq \alpha\leq 2$ is called the Levy index. 
In the following we will work in units of $r_0$ so that, without loss
of generality, we set $r_0=1$.
We emphasize that the step length is always larger than $r_0=1$ as is
more apparent on the the probability density function
\begin{align}
\label{eq:pdf}
  f(r)=\dfrac{\alpha}{r^{\alpha+1}} \Theta(r-1),
\end{align}
where $\Theta$ is the Heaviside step function.

This power-law form implies infinite variance
and also in some cases ($\alpha\leq1$) infinite expectation value. 
The random walk, i.e. the cumulative sum of independent steps
distributed this way,  
fail the \CLT and does \textit{not} converge to a Gaussian distribution.
Instead, following the work of P. Levy \cite{Levy:1925,Levy:1954}, it was shown that it has a
basin of attraction to what is known as $\alpha-$stable distributions
\citeg{Paul:2013}. This is called the Generalized Central Limit
Theorem \cite{Khintchine:1938,Gnedenko:1954} (the Gaussian case is recovered for $\alpha=2$).

A Levy flight can be simulated in a space of any dimension by
drawing isotropically a direction, and a radial part following 
\refeq{LF}.
We will mostly focus our attention on dimension 2, since it is the
dimension for which the model best describes the data. 
In this case, the probability to return
close to a previous point of the walk, which we loosely call the
``return-probability'' is most important.

The Levy flight trajectory has a fractal dimension $\alpha$ \cite{Seshadri:1982}; the mean
density of points within a sphere of radius $R$ localized on one of the point
of the flight, varies as $1/R^{2-\alpha}$
\cite{Mandelbrot:1975,Plaszczynski:2022}. 
The process is statistically \textit{self-similar}; loosely speaking
the pattern of points  ``looks the same'' at any scale. 
A more precise description is given in \ref{app}.
We will see that, due to
this self-similarity, several power-law distributions (scaling
relations) emerge.
An introduction to the (rich) theory of Levy flights is given in \cite{Chechkin:2008}.

\begin{figure}
  \centering
  \subfigure[]{\includegraphics[width=\linewidth]{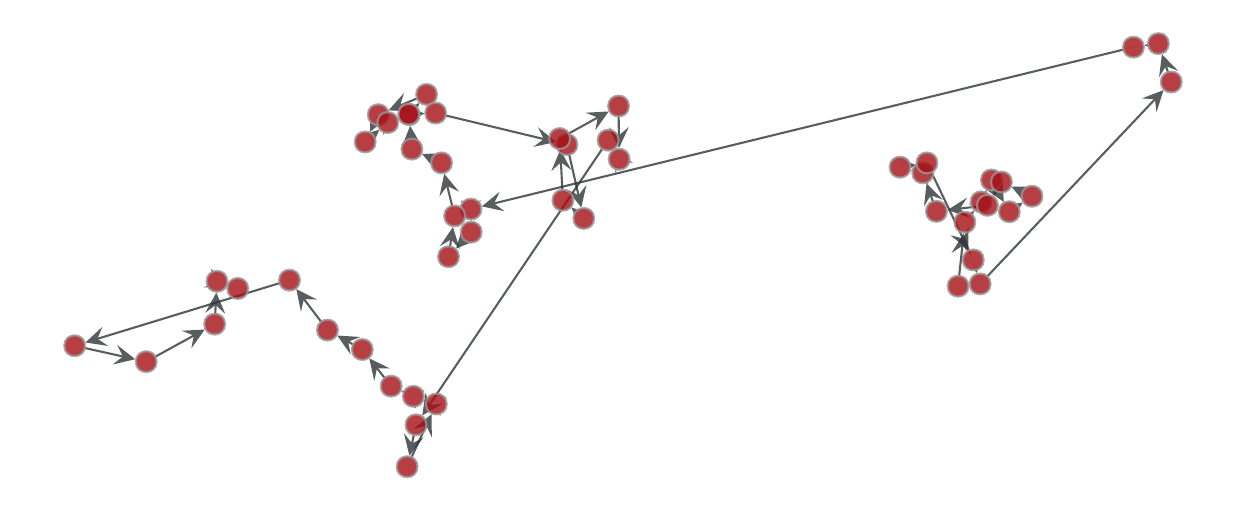}}
  \subfigure[]{\includegraphics[width=\linewidth]{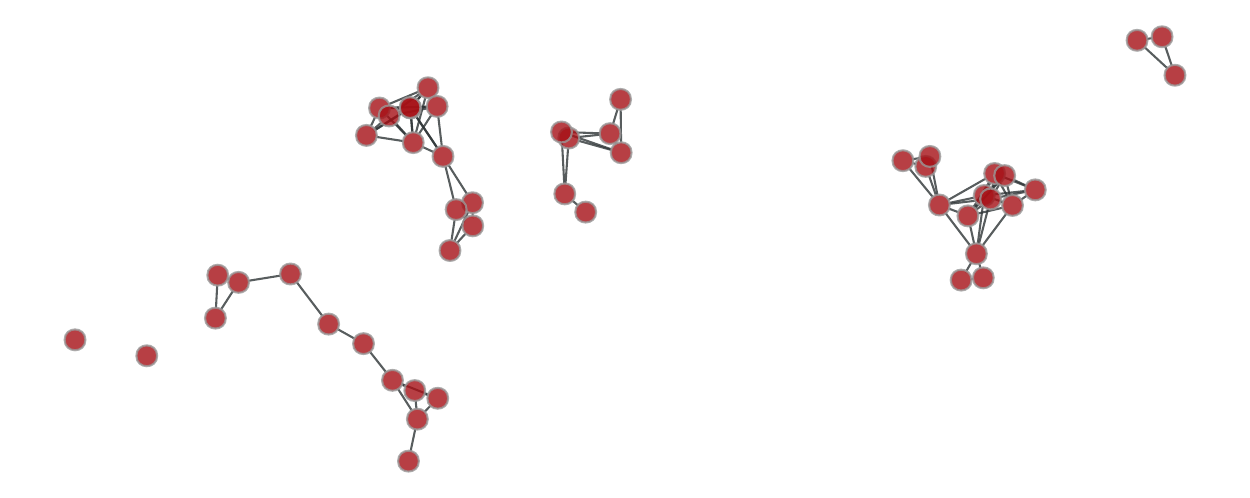}}
  \subfigure[]{\includegraphics[width=\linewidth]{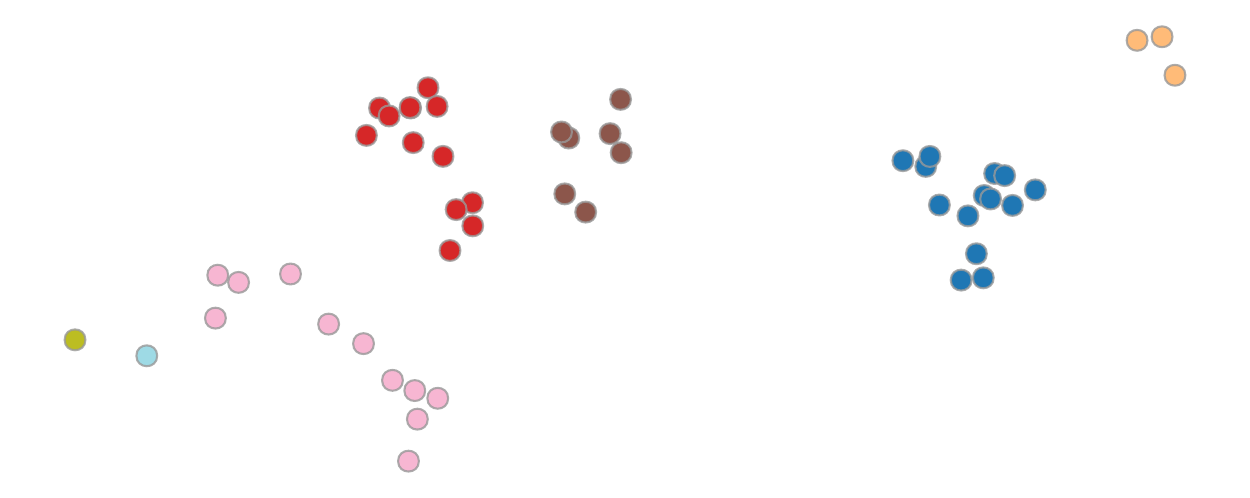}}
  \caption{Example of construction of Levy clusters. (a) Levy flight.
    The points are obtained from a random Levy walk of index $\alpha=1.1$
    and the line shows their creation path. (b) Levy graph. The
    points form the vertices of a graph with edges connecting pairs
    if their distance is inferior to the scale $s$ (here $s=3$). (c) The connected components
    of the graph form Levy clusters which are colored the same. In our 
    model, their size represent the duration of contact
    between two individuals and the scale is related to their interaction mean-time.}
  \label{fig:levy}
\end{figure}

A distinct feature of Levy flights is their clustering properties.
Several points of the walk end nearby before a long jump triggers the
construction of another cluster. Because of the absence of scale in
\refeq{LF} a nested hierarchy of self-similar clusters is created \cite{Hughes:1981}.
To study quantitatively the clusters' properties we need however to
apply some scale. This can be performed by joining pairs of points which are below some distance $s$ to form the edges
of a geometric graph. The effect of applying explicitly a scale onto
a scale-free process was studied in \cite{Plaszczynski:2022} and 
lead to a family of graphs $\lgg{\alpha}{s}$, called 
``Levy Geometric Graphs''.
They consist in a set of connected-components that we call
(Levy) ``clusters''. An example is show in \Fig{levy}.
Although not (yet) justified, it was noticed that the mean 
number of clusters for a Levy graph of size $N$ scales as
\begin{align}
\label{eq:nclus}
 \mean{\Nclus}/N=A_c/s^{\alpha_c},
\end{align}
where $A_c\to1,~\alpha_c\to\alpha$ in a large dimension space, and are 
obtained from numerical simulations otherwise (\Sup{S2}).

\section{An empirical model of discussion}
\label{sec:data}

An intriguing fact is that the cluster size contrast for Levy graphs  (Fig.8 of \cite{Plaszczynski:2022}), i.e.
$n_i/\bar n$ where $n_i$ denotes the size of the connected components,
has a distribution that looks 
similar to what is observed for the duration of contacts in the
\textit{sociopatterns} data (Fig.4 of \cite{Plaszczynski:2024}).
We thus try to match a model based
on Levy graphs to the description of the observed duration contrast.
To this purpose, on each \textit{sociopatterns} dataset, we extract
the set of all relations
$r$ and compute their total $w(r)$ and mean $\bar t(r)$ interaction time.
Then, for each relation we generate a 2D Levy graph of size $N=w(r)$
\footnote{Since we use raw data, $w(r)$ is expressed in units
of the resolution step and has thus an integer value.}.
To match the observed mean-time to the mean cluster size we use \refeq{nclus}
\begin{align}
   \mean{\Nclus}/N=1/\bar n =A_c/s^{\alpha_c}=1/\bar t,
\end{align}
which sets the scale to
\begin{align}
\label{eq:st}
  s(r)=\left[A_c \bar t(r)\right]^{1/\alpha_c},
\end{align}
where the $A_c(\alpha)$ and $\alpha_c(\alpha)$ values are taken from \Sup{S2}.
We then extract the connected-components of the graph and measure
their size ($n_i$) and
compute the cluster size contrast $\delta_i=n_i/\bar n$ distribution.
Since the graph is random, we repeat the procedure 10 times and superimpose the results.
The only free parameter in the model is the Levy index $\alpha$.

\begin{figure}[htbp!]
  \centering
  \includegraphics[width=\linewidth]{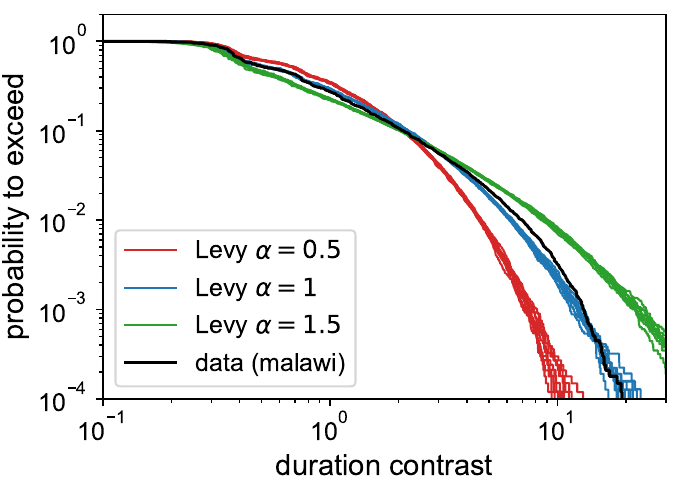}
  \includegraphics[width=\linewidth]{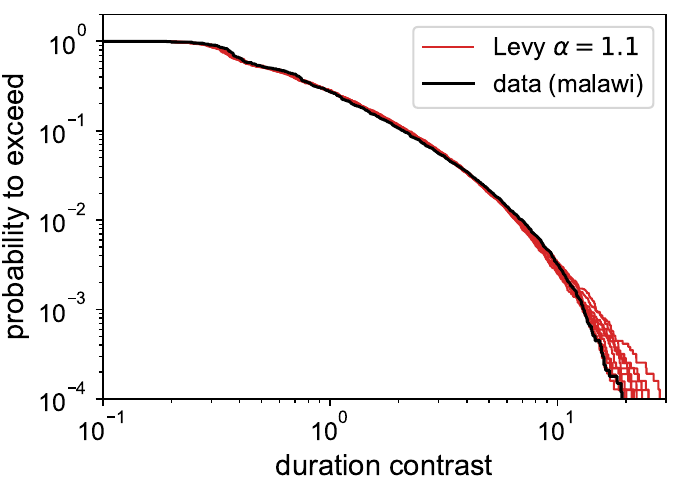}
  \caption{Upper: Comparison between the cluster size distribution contrast
    (\ccdf) of the Levy model 
    defined in the text for 3 indices ($\alpha=0.5,1,1.5$) and the contact duration
    contrast measured from the \textit{sociopatterns} data recorded
    in a Malawi village \cite{Ozella:2021}. Each line of the same color represents a
    different realization of the random graph. Lower: the best
    agreement is observed for $\alpha=1.1$.}
  \label{fig:malawimodels}
\end{figure}

We show how the distributions vary for some $\alpha$ values on the
upper part of \Fig{malawimodels}.
There is little spread among the different realizations and, as shown
in the bottom part of \Fig{malawimodels}, a value around
$\alpha=1.1$ reproduces well the data 
(a more precise determination will be given later).

\begin{figure}[htbp!]
  \centering
  \includegraphics[width=\linewidth]{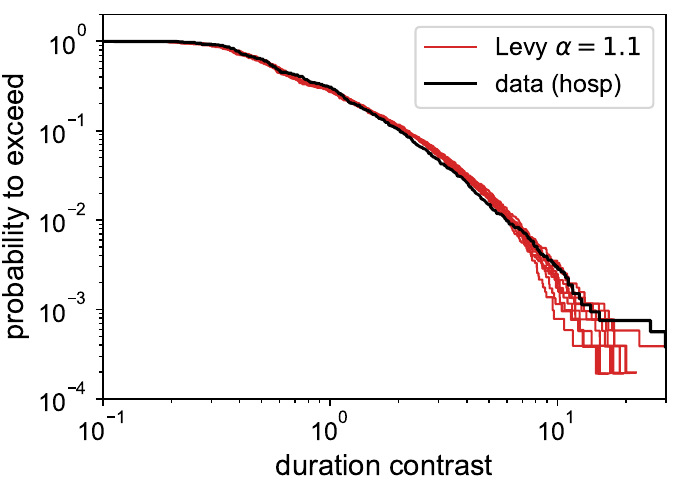}
  \includegraphics[width=\linewidth]{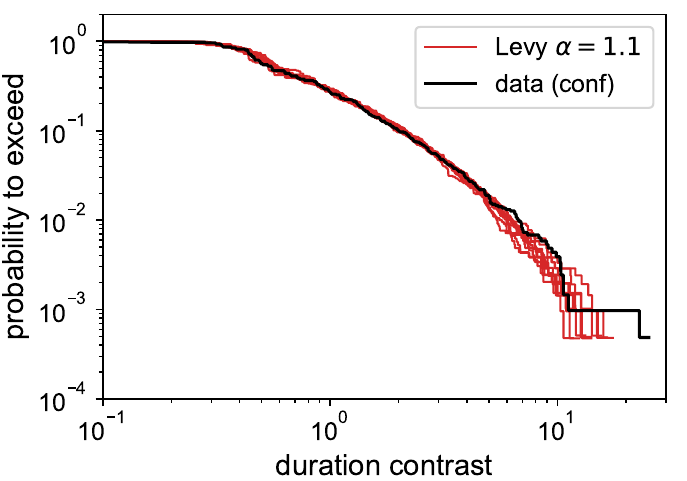}
  \includegraphics[width=\linewidth]{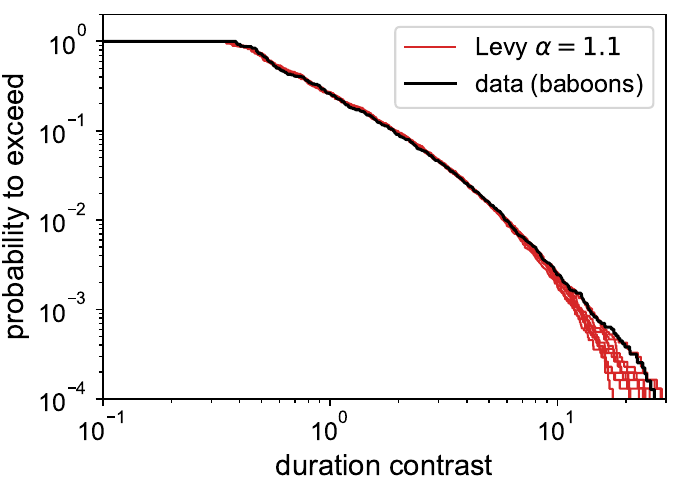}
  \caption{Comparison between the cluster size distribution contrast of the Levy model 
    with a $\alpha=1.1$ index and the contact duration
    contrast measured on several very different \textit{sociopatterns}
    datasets. Upper: in a french hospital, middle: at an international
    conference, lower: between baboons in an enclosure.}
  \label{fig:othermodels}
\end{figure}

We then show in \Fig{othermodels} that the same model
accommodates also perfectly three other \textit{sociopatterns} datasets while they
were taken in some very different sociological environments (see
\Sup{S1} for details).
It is remarkable that such a level of agreement is met on all the
datasets \textit{with 
the very same index} (which is the only parameter in the model).
Even the tiny fluctuations
at low contrast are well reproduced. While the fact that the data
distributions are similar is already known (it is the main topic of
\cite{Plaszczynski:2024}), the Levy graphs are
built from very different set of relations and characteristics 
(interaction rates and mean-time). Thus the graphs' size and scale vary
greatly between the datasets and it is not a trivial result that
the same index can match the different datasets nor that the low-contrast
part can be reproduced so well.

\section{Justification of the model}
\label{sec:dynamics}

Although the agreement of the model with the data is impressive, 
it clearly lacks justification. To this purpose we need to understand how
clusters form during the evolution of the Levy flight, i.e.
the dynamic of clusters formation in this peculiar form of random-walk.
We proceed thus to computing analytically the distribution of the
clusters size for any Levy geometric graph which will be shown to be
of stretched exponential type.

\subsection{The dynamics of Levy clusters formation}

\begin{figure}[h!]
  \centering
  \includegraphics[width=\linewidth]{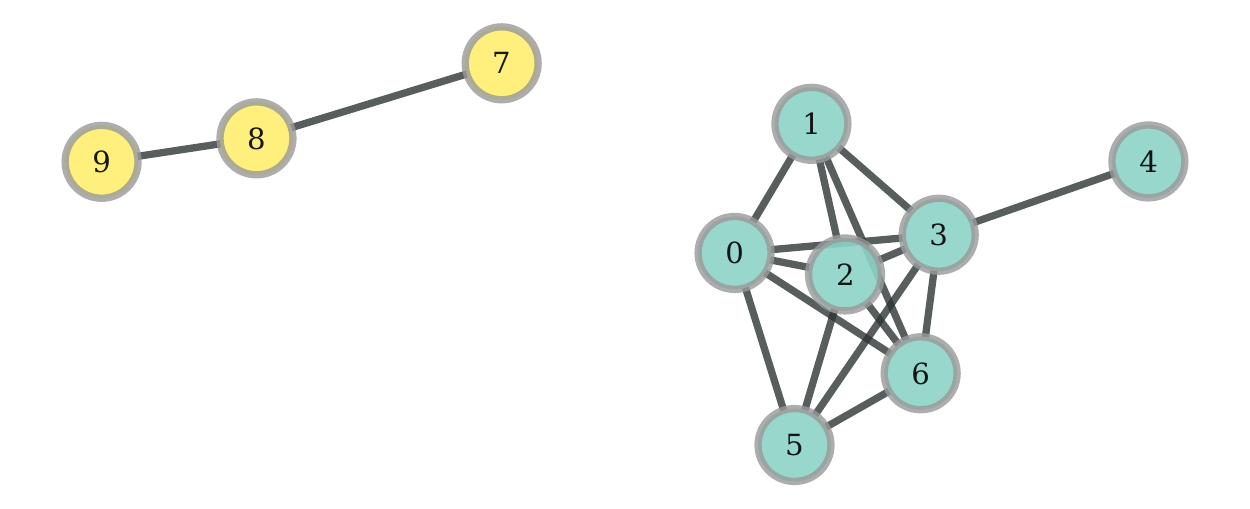}
  \caption{Example of two clusters obtained on a \lgg{1}{3}
    graph. The numbers indicate the order of appearance of the points
    during the flight. Points are colored according to the cluster
    they belong to. The edges show pairs connected when their distance
    is below $s=3$.
}
  \label{fig:clusters}
\end{figure}

We consider how clusters grow in a Levy graph when new points enter
the flight. \Fig{clusters} shows some typical cases for the clusters growth.

\begin{itemize}
\item The yellow cluster consists of three points (7,8,9), each
  being connected to the previous one.  
This is a  case where the  distances between the consecutive steps
all fall below the chosen scale (here $s=3$). For reasons that will
become clear later we call this the \textit{geometric configuration}.
\item The light-blue cluster (0,1,2,3,4,5,6) shows a more evolved
  configuration. Points are linked to their predecessors but for the
  (4-5) pair where the distance exceeds the scale. Point (5) is 
  however close enough to (0), and later to (6), to be incorporated into
  the cluster. This effect is due to what we call the \textit{return} (or
  \textit{trapping}) \textit{probability} which increases when there are more points.
  It is a ``rich-get-richer'' mechanism which will be investigated later.
\end{itemize}
Although both aspects are intimately related, we consider them first 
separately in order to understand their properties.

\subsubsection{The geometric case}
\label{sec:bernouilli}
\begin{figure}[h!]
  \centering
  \includegraphics[width=\linewidth]{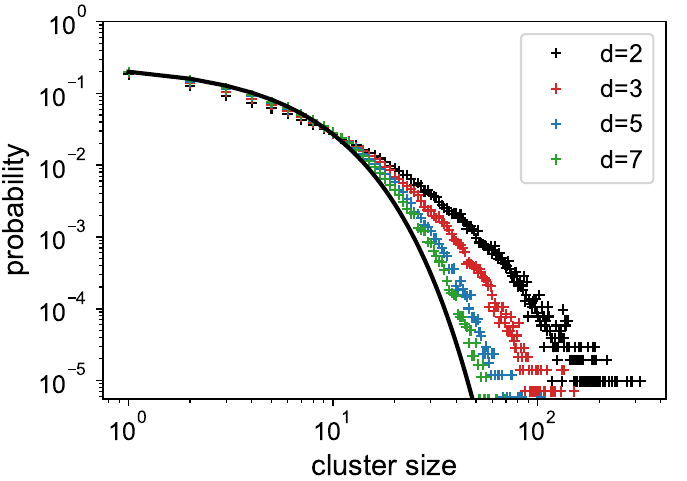}
  \caption{Distribution of the clusters size for a \lgg{1}{5} graph
    obtained by simulation ($N=10^6$ steps) increasing the space
    dimensionality $d$ (points). The distributions converge to the
    geometric form (black line).}
  \label{fig:bernoulli}
\end{figure}

Let us imagine that we can switch off the angular part in the
jumps keeping only the radial distribution \refeq{LF}. This is in fact
possible, by performing a Levy flight on a 1D line but always in the same
direction. One can approach also this regime by 
increasing the dimensionality of the space. Indeed in many dimensions 
it becomes very unlikely that a new jump connects to a
previous point of the walk since it will most often go into other parts of the
space.
Let us call a ``success'' a jump that exceeds the distance $s$.
Without a return probability, the only way to form a cluster is by
linking the new point to the previous one.
To form a cluster of size $n$, one then needs $n-1$ consecutive
``failures''.
The corresponding probability is the geometric distribution
\begin{align}
\label{eq:bern}
  \bern{n}=p_0(1-p_0)^{n-1},
\end{align}
where $p_0$ is the probability to escape the cluster, which, according to
\refeq{LF}, 
is
\begin{align}
  \label{eq:p0}
  p_0=1/s^\alpha,
\end{align}

and $n$ starts at 1 (the minimal cluster size). The expectation value for the size of a cluster is 
\begin{align}
\label{eq:meanbern}
  \E{n}=1/p_0=s^\alpha,
\end{align}
in agreement with \refeq{nclus} for a large dimension space, since the mean number of
clusters in a graph of size $N$ is $\mean{\Nclus}=N/\mean{n}$.
We show in \Fig{bernoulli}, using simulations, that when increasing the dimension of the
space for the Levy flight, the cluster size distribution 
converges indeed to the geometric one.

\begin{figure}[htbp]
  \centering
  \includegraphics[width=\linewidth]{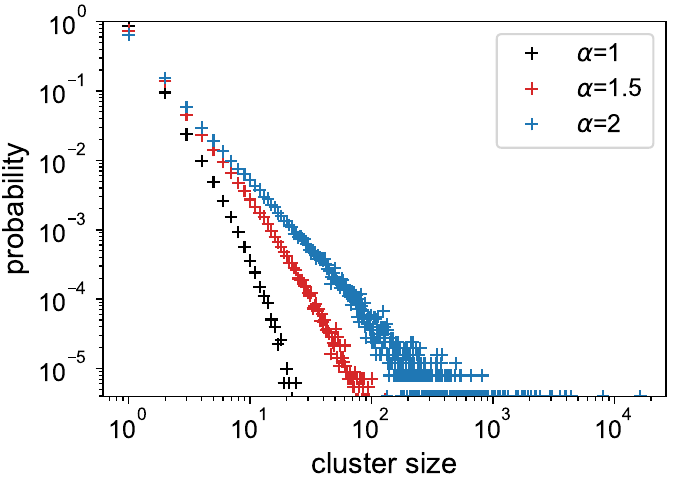}
  \caption{Distribution of 2D Levy clusters size with a scale $s=1$
    for $\alpha=1,1.5$ and 2.}
  \label{fig:s1}
\end{figure}

We now consider the walk in dimension 2 where the effect of the return-probability cannot be
neglected especially for large clusters (\Fig{bernoulli}). To gain intuition on its effect on cluster formation we
would like to cancel the geometric configuration. This may be performed in
the following way.
Remembering that a single Pareto-Levy step always produces a point at a
distance beyond $r_0=1$ (\refeq{pdf}), if we construct a Levy graph at a
scale $s=1$ or below, two
consecutive points can never get connected. Without a return
probability, this would lead only to a set of singlet clusters. 
However the return-probability brings back some points from the
walk close to previous ones opening the possibility to create Levy
clusters with more than a single vertex.
\Fig{s1} shows the resulting distributions obtained in simulations
with a scale $s=1$ and varying the Levy index. They fall steeply 
and converge to a power-law as $\alpha$ reaches 2, which will be
confirmed analytically later.
The emergence of heavy tails that are close to power-law ones is
typical of a preferential-attachment mechanism that we shall now investigate.

\subsubsection{Preferential attachment}
\label{sec:rgr}

As presented in the introduction, discussions benefit from a
preferential attachment effect:  
when they last for long it is likely they will
extend even more. It becomes more and more difficult to escape a
conversation while it lasts.
This feature is captured in our model by the Levy clusters.
When the size of a cluster increases, its trapping probability
increases too; 
it gets more and more difficult to \textit{escape the cluster}, i.e. fall in
a region of the space far from the cluster points.
We may get an order of magnitude of this effect with the following argument.
In 2D in average, $k$ points in a cluster occupy an area $\pi R_c^2$, so that
the mean cluster radius varies as $R_c\propto\sqrt{k}$.
The probability to escape the cluster with a single step, i.e. perform
a step beyond $R_c$ is then of the order of (see \refeq{LF})
\begin{align}
\label{eq:a2}
  p(r>R_c)\propto\dfrac{1}{k^{\alpha/2}}.
\end{align}

While $k$ increases the success probability (to escape the cluster)
decreases. A cluster of size $n$ occurs when there are $n$ consecutive failures.
We thus generalize the geometric form \refeq{bern} to
\begin{align}
\label{eq:pn}
 p(n)\propto {\prod_{k=1}^{n}(1-p_0(k))},
\end{align}
where
\begin{align}
  p_0(k)=\dfrac{A}{k^\gamma},
\end{align}
and $\gamma$ should be close $\alpha/2$. 
The normalization is given by $\sum_n p(n)=1$ and 
the geometric configuration is recovered by canceling the preferential attachment
term with $\gamma\to0$ and $A\to p_0=1/s^\alpha$.

\begin{figure}[htbp]
  \centering
  \includegraphics[width=\linewidth]{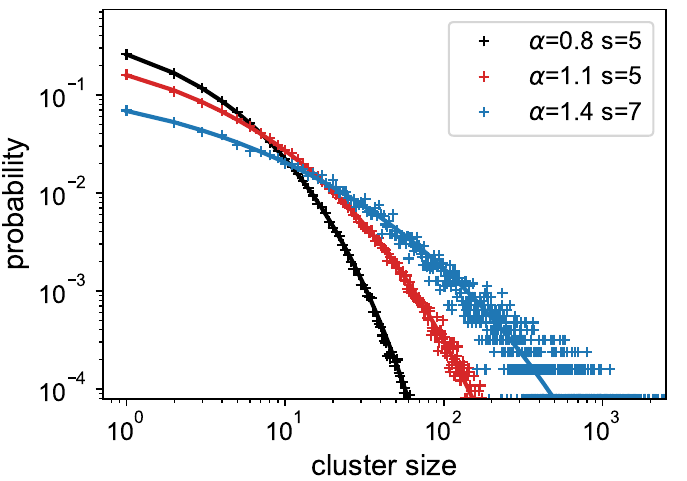}
  \caption{ Clusters size distribution obtained from Levy graphs
    simulations for some $\alpha$ and $s$ values (points).
    The curves show the corresponding \refeq{pn} model, adjusting the
    $A$ and $\gamma$ parameters with least-squares minimization.}
  \label{fig:fits}
\end{figure}

We confront this formula to the measured cluster sizes obtained in 
simulations, performing a least-squares
minimization on $A$ and $\gamma$, for several values of $\alpha$ and $s$ varied in the range
$0.8\leq\alpha\leq1.4$ and $3\leq s \leq 10$.
We obtain an excellent agreement in each case as shown on some
examples in \Fig{fits}.

\begin{figure}[htbp]
  \centering
  \includegraphics[width=\linewidth]{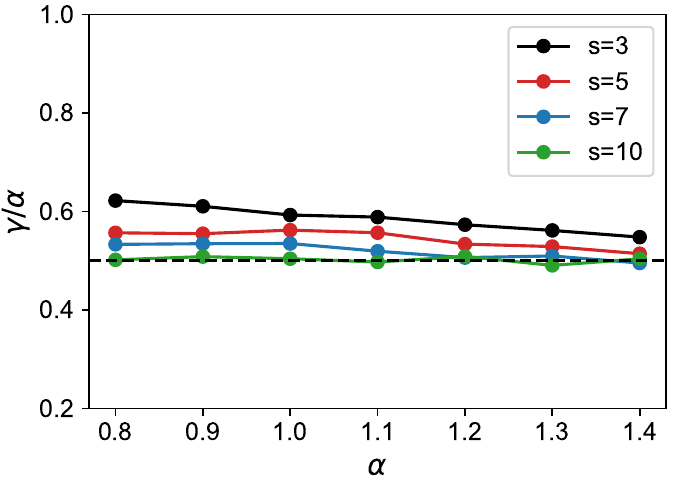}
  \caption{Ratio between the $\gamma$ parameter of \refeq{pn} and
    $\alpha$ determined from \lgg{\alpha}{s} simulations varying
    $\alpha$ for several scales $s$. The dashed line at 0.5
    represents the \refeq{a2} expectation.}
  \label{fig:gamma_a}
\end{figure}

We now consider the values of $\gamma$ and $A$ extracted from the fits.
As shown in \Fig{gamma_a} the exponent $\gamma$ depends essentially
on $\alpha$ and at large scales (corresponding to large $R_c$ in
\refeq{a2}) converges indeed to $\alpha/2$.

\begin{figure}[ht!]
  \centering
  \includegraphics[width=\linewidth]{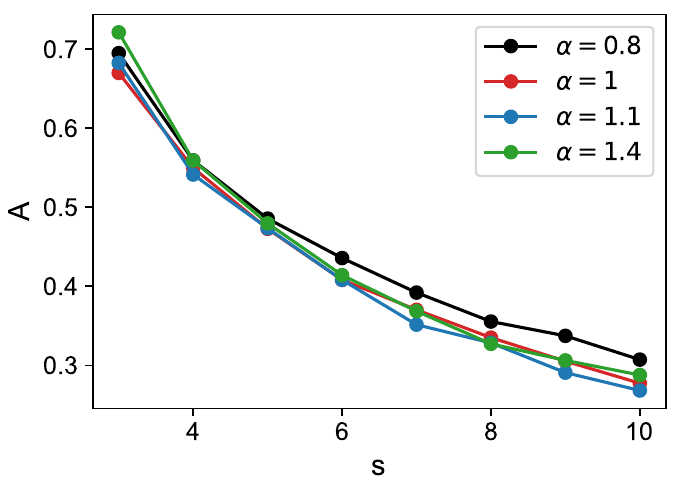}
  \caption{Value of the amplitude term $A$ in \refeq{pn} 
    determined from simulations as a function of
    the scale $s$ for several $\alpha$ values.}
  \label{fig:A_s}
\end{figure}

On the contrary, the amplitude term $A$ mostly depends on the scale as
shown in \Fig{A_s} with a power-law dependency of the form
\begin{align}
\label{eq:beta}
  A(s)=c/s^\beta.
\end{align}
The $c$ and $\beta$ coefficients are obtained from simulations (\Sup{3}) and for
$\alpha$ around 1 are described by
\begin{align}
  c&=1.19+0.31\alpha \\
  \beta&=0.57+0.14\alpha,
\end{align}
with a $\pm0.02$ precision on each parameter.
For the discussion model $c=1.5,\beta=0.7$.

\begin{figure}[ht!]
  \centering
  \includegraphics[width=\linewidth]{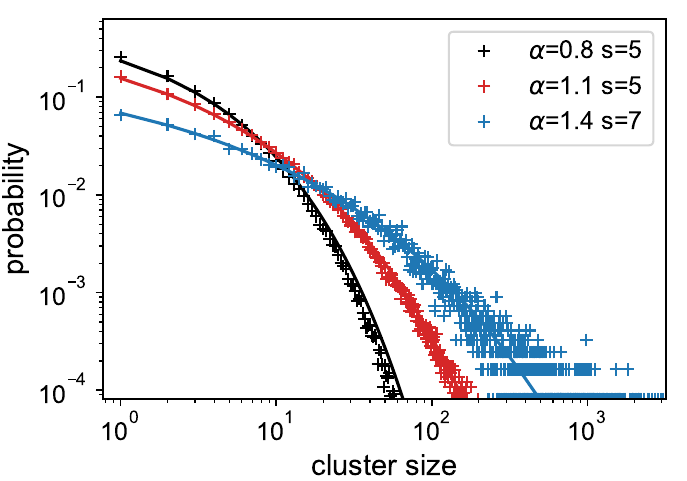}
  \caption{Distribution of the clusters size 
    (points) compared to the stretched exponential model
    \refeq{asym} derived in the text (lines).}
  \label{fig:asym11}
\end{figure}

We thus have the whole description of the clusters size probability
function
\begin{align}
 p(n;\alpha,s)\propto {\prod_{k=1}^{n}\left( 1-\dfrac{c}{s^\beta k^\gamma}\right)},
\end{align}
which may be further simplified.
Since 
\begin{align}
  p(n)/p(n-1)=1-A/n^\gamma,
\end{align}

the logarithmic slope goes asymptotically as
\begin{align}
  D=\dfrac{\ln{p(n)}-\ln{p(n-1)}}{\ln{n}-\ln{(n-1)}} \sim -A n^{1-\gamma},
\end{align}
with $\gamma\simeq\alpha/2$. 
For $\alpha=2$ we obtain a pure power-law form as was observed in \Fig{s1}.
Considering the continuum case, $D=\tfrac{n}{p}\tfrac{d p}{ dn}$, we
find that, for $\alpha<2$, the distributions follows asymptotically that of a stretched
exponential
\begin{align}
\label{eq:asym}
  p(n) \propto \exp{\left(-\tfrac{A}{1-\gamma}n^{1-\gamma}\right)}.
\end{align}
Numerically it appears that this approximation is in fact good on the whole
range of cluster sizes as shown in \Fig{asym11}.

The distribution for the discussion model is
\begin{align}
\label{eq:asym11}
  p(n;\alpha=1.1,s\gg1) \propto \exp{\left(-\dfrac{3.3}{s^{0.7}}n^{0.45}\right)}.
\end{align}

The generalization to a space of any dimension can be easily performed.
Adapting the argument leading to \refeq{a2} to a ball in a $d$ dimensional space
\begin{align}
  \gamma=\alpha/d.
\end{align}
When changing the dimension, the parameters $c$ and $\beta$ of the
amplitude term $A=c/s^\beta$ evolve but the asymptotic expression
\refeq{asym} still holds as shown in the \Sup{S4}.
When $d$  increases $\gamma\to0$ and $c\to1, \beta\to\alpha$ and 
the cluster size distribution converges to 
$\exp(-n/s^\alpha)$ which is the continuous extension of the geometric
function as was observed in \Fig{bernoulli}.

\subsection{Scaling relations}
\label{sec:scaling}

We may now compute the mean cluster size using the asymptotic
expression. With the shorthand notations
$\mu=1-\gamma,~\lambda=A/\mu=c/(\mu s^\beta)$
\begin{align}
  \bar n=\dfrac{u}{v}=\dfrac{\int_1^\infty n e^{-\lambda n^\mu} dn}{\int_1^\infty e^{-\lambda n^\mu} dn}.
\end{align}

With the change of variable $t=\lambda n^\mu$ the denominator becomes
\begin{align}
\label{eq:v}
v=\dfrac{\lambda^{-1/\mu}}{\mu}\Gamma(1/\mu,\lambda) ,
\end{align}
where the incomplete $\Gamma$ function is defined by
\begin{align}
\label{eq:gamma}
  \Gamma(a,z)=\int_z^\infty t^{a-1}e^{-t} dt. 
\end{align}
Integrating by parts and performing a similar computation on the
numerator yields 
\begin{align}
  u=\dfrac{\lambda^{-2/\mu}}{\mu}\Gamma(2/\mu,\lambda^2),
\end{align}
and the mean number of clusters finally reads
\begin{align}
\label{eq:nbar}
  \bar n=\dfrac{\lambda^{-1/\mu}\Gamma(2/\mu,\lambda^2)}{\Gamma(1/\mu,\lambda)}.
\end{align}

When the scale increases $\lambda=c/(\mu s^\beta)\to0$ and
asymptotically the mean cluster size scales with $s$ as 
\begin{align}
\label{eq:nbarpow}
\bar n=\dfrac{\Gamma(2/\mu)}{\Gamma(1/\mu)}
  \left(\dfrac{\mu}{c}\right)^{1/\mu} s^{\beta/\mu},   
\end{align}

\begin{figure}[ht!]
  \centering
  \includegraphics[width=\linewidth]{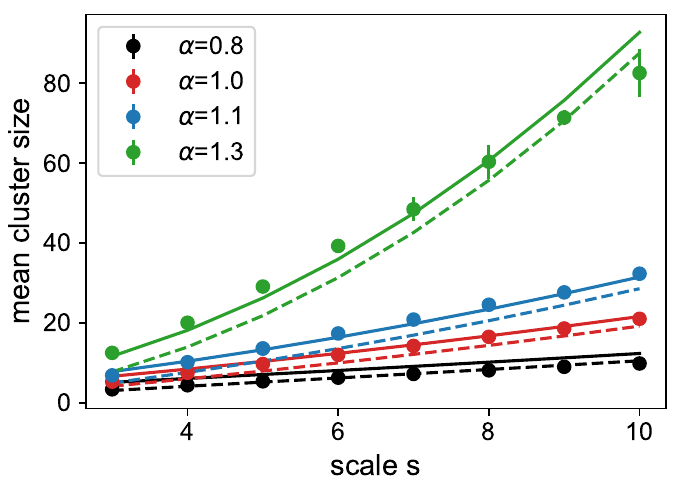}
  \caption{Mean clusters size of \lgg{\alpha}{s} graph's measured on
    the simulation (points) varying the scale for a few $\alpha$
    values, compared to the exact analytical model \refeq{nbar} (full
    lines) and its
    approximation \refeq{nbarpow} (dashed lines).}
  \label{fig:nbar}
\end{figure}

This regime is however only reached for very large scales ($s\gtrsim 10$).  Although
the $\Gamma(2/\mu,\lambda^2)$ function in the numerator of \refeq{nbar}
converges rapidly to $\Gamma(2/\mu)$, this is
not the case for $\Gamma(1/\mu,\lambda)$ in the denominator with $\beta\simeq0.7$. 
Numerically the variation of this term with $s$ in a limited scale
range is smooth and leads essentially to a constant correction term in
\refeq{nbarpow}.
This is shown in \Fig{nbar} where we compare the results from
simulations to \cref{eq:nbar,eq:nbarpow}. The agreement is
good given the uncertainties on the fitted $\gamma,c$ and
$\beta$  parameters.

Because of the long jumps in the Levy walk, the formation of a cluster
at a given scale is essentially a \textit{local} process. The walker
almost never ``comes back'' to a previous cluster after a long jump.
Clusters are thus mostly \textit{uncorrelated} and Levy graph's
may be viewed as a ``factory'' of clusters.
The sum of sizes of the \Nclus clusters equals the graph's size
\begin{align}
  \sum_{i=1}^{\Nclus} n_i=N.
\end{align}
In the mean, each (independent) cluster contributes to $\bar n$ so that
 $\mean{\Nclus} \bar n \simeq N$.
This explains the origin of the second scaling law on the fractional number of
clusters (\refeq{nclus})
\begin{align}
\label{eq:scale2}
 \mean{\Nclus}/N\simeq1/\bar n \propto 1/s^{\beta/\mu}.
\end{align}

\subsection{The cluster size contrast distribution}
\label{sec:contrast}

We now have all that is needed to compute analytically the distribution of the cluster
size contrast variable
$\delta=n/\bar n$, as observed in Figs.
\ref{fig:malawimodels} and \ref{fig:othermodels}. 
Performing the change of variable $n\to n/\bar n$ using
\cref{eq:asym,eq:nbarpow} and normalizing the result,  
we obtain
\begin{align}
\label{eq:pdfdelta}
  p(\delta) &= \dfrac{\mu\Gamma(2/\mu)}{\Gamma^2(1/\mu)}
  e^{- \left[\tfrac{\Gamma(2/\mu)}{\Gamma(1/\mu)} \delta\right]^\mu} 
 \quad \text{with}~ \mu=1-\dfrac{\alpha}{2}.
\end{align}

Integrating the formula, we obtain the \ccdf
\begin{align}
\label{eq:ccdf}
  p(\delta>x)& = \int_x^\infty p(\delta) d\delta \nn \\
             &  =\dfrac{\Gamma[1/\mu,\left(x \Gamma(2/\mu)/\Gamma(1/\mu)\right)^\mu]}{\Gamma(1/\mu)}
\end{align}
where the incomplete $\Gamma$ function is given by \refeq{gamma}.

For the discussion model we have 
\begin{align}
\label{eq:contrast11}
  p(\delta) & = 3.9~e^{-2.8 \delta^{0.45}}, \\
  p(\delta>x)&=0.9~\Gamma(2.2,2.8 x^{0.45}).
\end{align}

\begin{figure}[ht!]
  \centering
    \includegraphics[width=\linewidth]{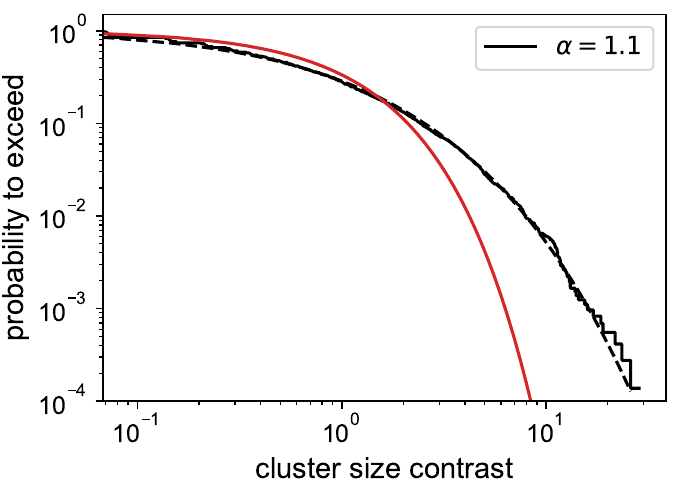}
  \caption{Distribution of the cluster size contrast for the
    discussion model based on a Levy graph of index 1.1. The full black line represent the \ccdf measured
    from simulations and the dashed one the analytical formula derived
  in the text (\refeq{contrast11}). The red curve corresponds to the
  geometric contribution.}
  \label{fig:contrast11}
\end{figure}

We show in \Fig{contrast11} the agreement of this formula with
simulations. We also added the contrast distribution of the Geometric
contribution which is \cite{Plaszczynski:2022}  

\begin{align}
  p(\delta>x)&=(1-p_0)^{x/p_0}.
\end{align}

It shows that short discussions are rather driven by the geometric
term while the tail arises from the preferential-attachment mechanism.

\section{The Levy index}

We can now determine the value of the Levy index on
the \textit{sociopattern} datasets. The analytical expression for the
contrast distribution, \refeq{pdfdelta}, must however be refined in order to build a precise likelihood
function by taking into account two effects.
\begin{itemize}
\item First, \refeq{pdfdelta} was derived for large
  scale values. Because of the low resolution of the
  \textit{sociopattern} data (T=20 s), the scales of the Levy graphs
  cover a rather low
  range ($1\lesssim s \lesssim 2$). Furthermore it was shown that the
  mean cluster size that is used in the computation (\refeq{nbarpow}) is slightly
  underestimated. To account for both effects we add a 
  normalizing factor $f$ that we compute from simulations so as to reproduce
  precisely the contrast distribution measured on Levy graphs . It is
  of the order of 1.3 depending on the dataset (details in \Sup{S5}).
\item Second, \refeq{pdfdelta} was derived asymptotically and
  since we are mostly interested in the tail of the distribution we 
  consider samples above $\delta_{min}=2$ (similar results are obtained
  with $\delta_{min}=3$). Care must then be taken on normalizing the 
  likelihood accordingly (as in \cite{Newman:2005}).
\end{itemize}

The improved contrast distribution reads
\begin{align}
  p(\delta;\alpha)&=\dfrac{g \mu}{\Gamma(1/\mu,\left(g
                    \delta_{min}\right)^{1/\mu})} e^{(-g\delta)^\mu}\\
  \text{with}~\mu=&1-\alpha/2,\quad g=f \frac{\Gamma(2/\mu)}{\Gamma(1/\mu)}.
\end{align}

The likelihood function for the set of measured values
$\delta_i>\delta_{min}$ is
\begin{align}
\label{eq:lkl}
  \mathscr{L}(\alpha)=\prod_{i} p(\delta_i,\alpha).
\end{align}

\begin{figure}[ht!]
  \centering
    \includegraphics[width=\linewidth]{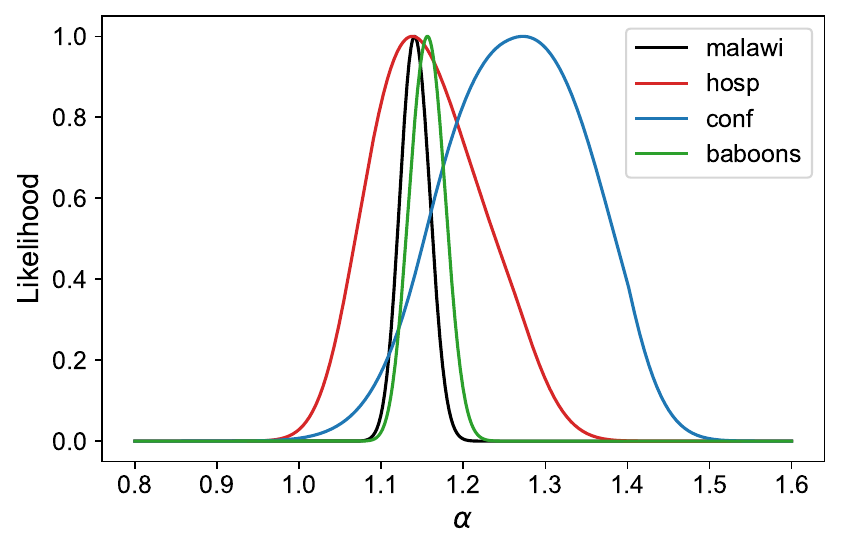}
  \caption{Likelihood functions \refeq{lkl} measured on the
    \textit{sociopattern} datasets.}
  \label{fig:likelihood}
\end{figure}

\Fig{likelihood} shows the normalized likelihood functions on the four datasets.

The maxima of these functions gives the maximum likelihood estimate of
the Levy index $\hat \alpha$ and the width of the support indicates the statistical precision of
the estimator.
To take also into account the stochastic nature of the model (based on
random graphs) we use simulations. For each dataset, we generate 100
realizations of the corresponding Levy graphs, compute each time the
maximum likelihood estimate and consider the spread of the resulting
distribution (see \Sup{S5}). This allows to obtain the standard
deviation of the estimator and its 95\% confidence level intervals.
Results are given in Table \ref{tab:MLE}.
The Levy indices determined on each dataset are compatible and
their combined value is $\hat \alpha=1.15\pm0.02$.

\begin{table}
  \centering
  \begin{tabular}{lccc}
    dataset & $\hat \alpha$ & $\sigma$ & 95\% CL\\
    \midrule
    \malawi & 1.14 & 0.02 & [1.11,1.18] \\
    \hosp & 1.14 & 0.06 & [1.07,1.26] \\
    \conf & 1.27 & 0.09 & [1.12,1.39] \\
    \baboons& 1.16 & 0.03 & [1.10,1.20]\\
    \bottomrule
  \end{tabular}

  \caption{Estimate of the Levy index for the different
    datasets. $\alpha$ is the Maximum Likelihood Estimate, $\sigma$
    its standard deviation and the last column indicates the 95\%
    confidence level interval. \label{tab:MLE}}
\end{table}

\section{Discussions as an aggregation process}
\label{sec:agg}

Our model somewhat resembles that of a growing network with a
sub-linear (power-law) kernel \cite{Krapivsky:2001}
In this model, nodes are added sequentially and attached to previous
ones with a probability related to their connectivity
$p(k)\propto k^\gamma$.
Asymptotically, the average number of sites with $k$ links grows
up linearly $N_k(t)=t n_k$ and  \cite{Krapivsky:2001}
\begin{align}
  \label{eq:GN}
  n_k&=\dfrac{\mu}{k^\gamma}
       \prod_{j=1}^k\left(1+\dfrac{\mu}{j^\gamma}\right)^{-1} .
\end{align}

There are some similarities with our model (\refeq{pn}) since for large $j$ values
$(1+\mu/j^\gamma)^{-1} \simeq 1-\mu/j^\gamma$.
There are however also a number of important differences.
\begin{itemize}
\item By construction, the network has a single connected-component
 and $k$ denotes the vertices degree.
\item In growing networks, there is a single degree of freedom, the
  amplitude term $\mu$ being related to $\gamma$ and $1<\mu(\gamma)<2$
  which is not what we measure for our amplitude term (\Fig{A_s}).
\item The fact that there is a single cluster gives rise
  to the $\mu/k^\gamma$ pre-factor which does not appear in \refeq{pn}.
\end{itemize}

\begin{figure}[h!]
  \centering
  \includegraphics[width=\linewidth]{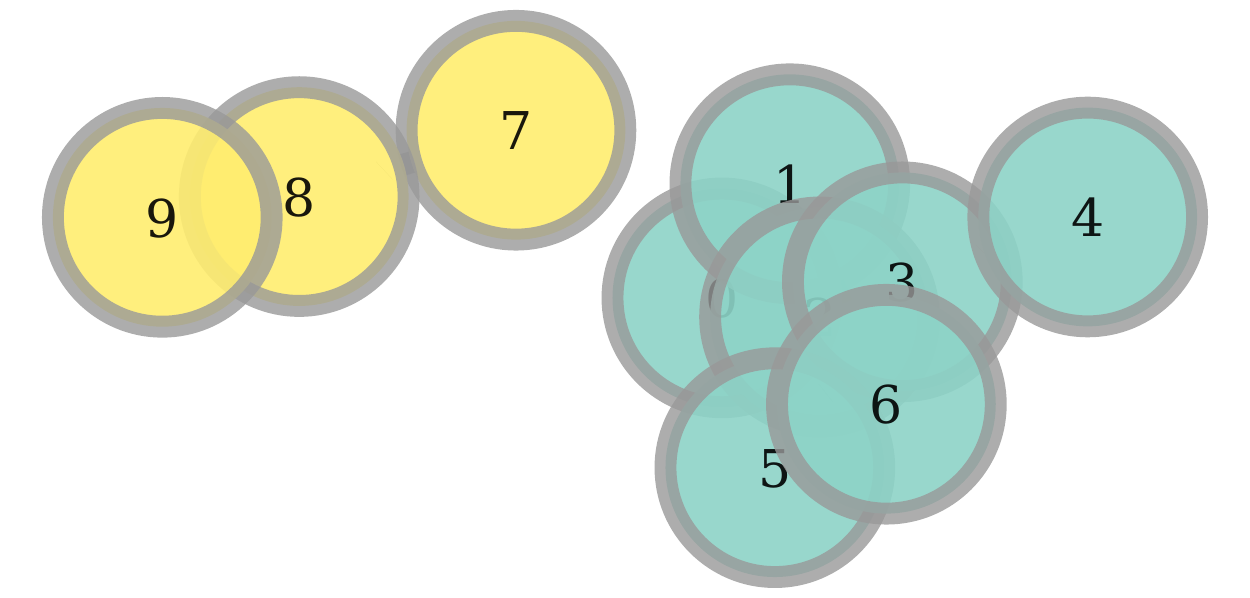}
  \caption{The formation of Levy clusters seen as an aggregation
    process. The Figure is the same as \Fig{clusters} when drawing disks
    of diameter $s$ around each vertex. Each discussion corresponds to
  a colored aggregate.}
  \label{fig:graphbig}
\end{figure}

The analogy with a growing network is thus limited although both
approaches lead to stretched exponential distributions.
The discussion model should be rather considered as an aggregation
process \footnote{The difference between both processes 
was already outlined in \cite{Krapivsky:2001}.}.
Two points get connected if their relative distance is below the
scale $s$. This is equivalent to considering some 
disks of diameter $s$ centered on each point that get connected if
they overlap  (see \Fig{graphbig}). Clusters/discussions between two
individuals thus appear as aggregates of such disks with a
natural radius determined by those persons mean-interaction time
(\refeq{st}).
As the aggregate (discussion) grows, 
it gets more likely for the next jumps of the disk to ``stick'' to it 
(conversation continues) which 
illustrates again the preferential attachment mechanism.

\section{Discussion(s)}
\label{sec:disc}

We have computed analytically the cluster size distribution for any
Levy geometric graph and found it to be of stretched-exponential type. 
But more importantly, we unraveled the
mechanism by which clusters form and its analogy with the way
discussions develop.

The initial phase is described by a geometric distribution, which for
a large scale is the exponential one.  For a fixed mean-value, 
it is the \textit{maximum entropy} distribution \cite{Jaynes:1957,Montroll:1983}. It means
that, beside its mean-time, there is no \textit{prior} information on the duration the conversation will take.

While more points are added to the cluster (discussion) it is more
likely that the next ones will be linked (related) since geometrically
the trapping probability increases.
This is a preferential-attachment mechanism and
reproduces the idea that as a discussion lasts it is more likely to
last even more.

In this model, discussions appears as an aggregation process between ``disks''
with a natural size fixed by the (usual) mean-time of the relation. 
Considering a discussion as a form of aggregation sounds reasonable.
Our main finding is that experimental data are very well reproduced when the disks
follow a 2D Levy-flight of index $\alpha=1.15$.

The duration of each discussion, characterized in the model by a number of points, is a \textit{discrete}
quantity expressed in units of the resolution step ($T=20 s$ for the
\textit{sociopatterns} data). Obviously we want a model that is independent from that resolution. 
With a better instrument (lower $T$) the duration and mean-time values
would increase.
In the Levy model this means changing the scale of the Levy graph
(see \refeq{st}) along with its number of points (size).
Among the class of random-walks only a Levy flight is resolution-independent. 
Indeed, when increasing the scale of any geometric graph based on a
standard (finite-variance) random-walk, we end-up with a single cluster,  
the giant component. 
When increasing the resolution, such a model would lead
to a single conversation between the two individuals. 
On the contrary, Levy-walk based graphs escape percolation \cite{Plaszczynski:2022}. 
Increasing the scale, the size of the graph changes and there are
still clusters. The important point is that the cluster size/duration
contrast $\delta=n/\bar n=t/\bar t$ is independent from the scale as
was derived in \refeq{pdfdelta}. 

A (measured) value of $\alpha=1.15$ is nothing special in the class of
Levy processes. Although there are some mathematical differences
when crossing the value $\alpha=1$ \cite{Metzler:2000} this value
looks excluded.
We note that several human activities exhibit a power-law
dependency with an exponent around 1 \citeg{Song:2010,Rybski:2009} but keep in mind that our
process is more elaborate than a mere power-law distribution. 
This could be a consequence of the ``least effort principle'',
as proposed by G. Zipf \cite{Zipf:1949} to explain several  
scaling exponents around 1 related to human behavior (such as the word
frequency), but a convincing explanation remains to be found.

The fact that the random walk is performed in a space of dimension 2 is
only dictated by the data. A walk in a higher dimensional space fits
the data less well (\Sup{S6}).

A ``point'' in our model denotes a time-unit.  It could
be related to a notion of information or more generally to a 
``topic'' (or concept). In the geometric
phase, which starts the discussion, each individual would discuss the 
initial topic, each one answering to the previous ``point''. The return-probability would
mean a break in this flow to a ``point'' previously raised. 
The preferential-attachment mechanism would mean that while more
topics have been addressed it is more likely to get back to one of them.
This is what is observed on internet discussion feeds
\cite{Delanoe:2015}. A person (``A'') posts some
question and gets an answer from ``B'', and there might be
an exchange between the two leading to a ``A-B-A...'' sequence. This
is what we identify as the geometric phase. While the discussion
goes on, third party users (``C'', then ``D'' etc.) may enter the
game, by commenting some part of the discussion which leads to
branches in the feed. The complexity of
the discussion increases, sometimes to a point where the whole
exchange gets out of control in solving the initial question. This
evolution looks somewhat similar to what we observe in
face-to-face discussions but there are some important differences too (there
are for instance several participants in the feed). Testing whether
such a hypothesis for the meaning of the word ``point'' makes sense 
requires the tools of Conversation Analysis \cite{Button:2022}.

\section*{Conclusion}

We have proposed a stochastic model to describe the duration of discussions between two
individuals. It reproduces precisely the distributions of the contrast
duration of contacts (i.e. the dimensionless deviation from the mean-value)
on several datasets of face-to-face interaction
recordings, with a single parameter, the Levy index, that was found 
to be $\alpha=1.15\pm0.02$.

The model is based on some simple considerations about conversations.
A discussion always starts about a topic that is first explored. While it lasts, it
becomes more likely to last even more since there are more points to come back to. 
Such a behavior can be reproduced by considering the size of the 
clusters (connected components) in geometric graphs built 
from the points of a Levy flight, named Levy Geometric Graphs \cite{Plaszczynski:2022}.
We derive analytically the cluster size distribution which is of
stretched-exponential type.
By dividing it by the mean-value to obtain the dimensionless ``duration
contrast'' variable, we show that it becomes
independent of the resolution. This is a necessary requirement for the
model which cannot be met by any other type of random-walk.
The analysis of the clusters dynamics reveals that they initially form
by linking each point to the previous one (geometric phase). Then as
more points are added more complex structure emerge 
due to the increasing return-probability to some previous points (preferential attachment).

We also show  that while the model shares some similarities with the case of
a growing network with a sub-linear kernel, it actually belongs to the class of
aggregation processes. The model may be
viewed as a ``disk'' with a diameter related to the mean-interaction time,
that makes jumps following a Levy flight of index 1.15 (see
\Fig{graphbig}). The size of each  set of overlapping disks represents the duration of a discussion.

Finally we may question the very notion of discussion and language. We have
disregarded the fact that baboon's interaction shows dynamics similar
to human one \cite{Plaszczynski:2024} and are thus also characterized in our model by a Levy index
of $\alpha=1.1$ (\Fig{othermodels}).
Although our communication through an evolved verbal language
is (probably) more complex than the baboons' one, this calls into question the very
notion of ``discussion'' and the role of language on the development
of the human species \cite{Heesen:2022}.

\appendix

\section{Levy flights self-similarity}
\label{app}

To understand the origin of the scaling relations, we come back to the
individual points of the Levy walk, that we label with time and
consider as complex-valued, $X(t)$.
Although the variance is formally diverging, one may get a similar
quantity by rescaling the lower-order fractional moments \cite{Metzler:2000}
\begin{align}
\label{eq:fracmom}
  \mean{\norm{X(t)}^q} & \propto  t^{q/\alpha}, \nn \\
  \mean{\norm{X(t)}^q}^{2/q} & \propto  t^{2/\alpha},
\end{align}
where $0<q<2$ for convergence.
This pseudo-mean squared displacement 
reveals the super-diffusive nature of the walk, with a Hurst exponent \citeg{Peitgen:2004}
\begin{align}
\label{eq:H}
H=1/\alpha.
\end{align}
This result may also be obtained on finite size samples
\cite{Bouchaud:1990} or using a sliding window during the walk
\cite{Jespersen:1999} 
or more rigorously with a Renormalization Group
approach \cite{West:1994}.
Self-similarity corresponds to the fact that $X(t)$ and
$X(rt)/r^{H}$ are statistically indistinguishable for any $r$ value,
i.e. that the samples follow exactly the same underlying distribution
(we illustrate it with simulations in \Sup{S7}).
We note note it $X(t)\equiv X(rt)/r^{H}$.
Taking $r=1/s^\alpha$ and using \refeq{H}, we find that 
\begin{align}
  X(t)/s\equiv X(t/s^\alpha),
\end{align}
This relation shows that \textit{rescaling} the random function random $Y(t)=X(t)/s$
is equivalent to \textit{sub-sampling} the original process by a factor
$1/s^\alpha$. This is in line with the definition of the effective diffusion
coefficient for a Weierstrass random walk \cite{Hughes:1981}, the analog of a
Levy flight but on a square lattice of mesh size $a$, given by
\citeg{Paul:2013}  $D_{eff}=\lim
_{a,\Delta t \to 0} a^\alpha/\Delta t$.

A Levy graph of scale $s$ is constructed by linking points satisfying
$\norm{X(t_i)-X(t_j)}<s$ or using the rescaled variable
$\norm{Y(t_i)-Y(t_j)}<1$.
Changing the scale is equivalent to varying the graph's size by
$N/s^\alpha$ but otherwise keeping topologically equivalent configurations. 
Then  all statistics related to the graph's size, vary 
with $s^\alpha$. For instance, the faction of  edges, average degree
or mean cluster size all scale with $s^\alpha$. The number of
clusters, being the inverse of the mean cluster size, \refeq{scale2}, scales inversely.

So why is the exponent we computed for the mean
cluster size (\refeq{nbarpow}) not exactly $\alpha$ but a more complicated
function of it? The reason lies in what we call the ``Levy
flight''. We have focused on the original definition
given by Mandelbrot \cite{Mandelbrot:1975}, not on the one used to derive \refeq{fracmom}
which is based on steps following \textit{exactly} a $\alpha$-stable
distributions (as a Cauchy one for $\alpha=1$).
The Pareto-Levy distribution used throughout the text (\refeq{pdf}) is not
exactly a stable one, because of the gap below $r_0=1$. However,
because of the Generalized CLT \cite{Gnedenko:1954}, 
its cumulative sum (i.e. the random-walk) converges rapidly to it.
While for most observables it has very little impact, it slightly
affects the clusters size (which are often made up from the very first
points of the walk) which makes the calculation of their distribution
more elaborate.

\section*{Acknowledgments}
We thank Alexandre Delano\"e for enlightening discussions on
conversation analysis and the analogy with internet discussion feeds.

\section*{Competing interests}
The author declare no competing interests.

\section*{Funding}
Public research.


\section*{Availability of data and material}
\begin{itemize}
\item The datasets analyzed in this study are available from the \textsl{sociopatterns} repository,  \safeurl{www.sociopatterns.org}
\item The \textsf{python3} software used to produce the results is available from 
\safeurl{https://gitlab.in2p3.fr/plaszczy/coll}
\item  All graph-related computations and figures were obtained with 
the \textsf{graph-tool} (\textsf{v 2.43}) software \cite{Graph-tool:2014}. 
\end{itemize}

\bibliographystyle{elsarticle-num}

\bibliography{references}

\begin{thebibliography}{10}
\expandafter\ifx\csname url\endcsname\relax
  \def\url#1{\texttt{#1}}\fi
\expandafter\ifx\csname urlprefix\endcsname\relax\def\urlprefix{URL }\fi
\expandafter\ifx\csname href\endcsname\relax
  \def\href#1#2{#2} \def\path#1{#1}\fi

\bibitem{holme:2012}
P.~Holme, J.~Saramäki, \href{http://arxiv.org/abs/1108.1780}{Temporal
  {Networks}}, Physics Reports 519~(3) (2012) 97--125, arXiv: 1108.1780.
\newblock \href {https://doi.org/10.1016/j.physrep.2012.03.001}
  {\path{doi:10.1016/j.physrep.2012.03.001}}.
\newline\urlprefix\url{http://arxiv.org/abs/1108.1780}

\bibitem{holme:2015}
P.~Holme, \href{http://link.springer.com/10.1140/epjb/e2015-60657-4}{Modern
  temporal network theory: a colloquium}, Eur. Phys. J. B 88~(9) (2015) 234.
\newblock \href {https://doi.org/10.1140/epjb/e2015-60657-4}
  {\path{doi:10.1140/epjb/e2015-60657-4}}.
\newline\urlprefix\url{http://link.springer.com/10.1140/epjb/e2015-60657-4}

\bibitem{Button:2022}
G.~Button, M.~Lynch, W.~Sharrock, Ethnomethodology, {{Conversation Analysis}}
  and {{Constructive Analysis}}: {{On Formal Structures}} of {{Practical
  Action}}, 1st Edition, {Routledge}, {London}, 2022.
\newblock \href {https://doi.org/10.4324/9781003220794}
  {\path{doi:10.4324/9781003220794}}.

\bibitem{Plaszczynski:2024}
S.~Plaszczynski, G.~Nakamura, B.~Grammaticos, M.~Badoual,
  \href{http://dx.doi.org/10.1140/epjds/s13688-023-00444-z}{On the duration of
  face-to-face contacts}, EPJ Data Science 13~(1) (Jan. 2024).
\newblock \href {https://doi.org/10.1140/epjds/s13688-023-00444-z}
  {\path{doi:10.1140/epjds/s13688-023-00444-z}}.
\newline\urlprefix\url{http://dx.doi.org/10.1140/epjds/s13688-023-00444-z}

\bibitem{Barabasi:1999}
R.~Albert, A.-L. Barab\'asi,
  \href{http://dx.doi.org/10.1126/science.286.5439.509}{Emergence of {Scaling}
  in {Random Networks}}, Science 286~(5439) (1999) 509–512.
\newblock \href {https://doi.org/10.1126/science.286.5439.509}
  {\path{doi:10.1126/science.286.5439.509}}.
\newline\urlprefix\url{http://dx.doi.org/10.1126/science.286.5439.509}

\bibitem{Barabasi:1999b}
R.~Albert, A.-L. Barab\'asi, H.~Jeong,
  \href{http://dx.doi.org/10.1016/S0378-4371(99)00291-5}{Mean-field theory for
  scale-free random networks}, Physica A: Statistical Mechanics and its
  Applications 272~(1–2) (1999) 173–187.
\newblock \href {https://doi.org/10.1016/s0378-4371(99)00291-5}
  {\path{doi:10.1016/s0378-4371(99)00291-5}}.
\newline\urlprefix\url{http://dx.doi.org/10.1016/S0378-4371(99)00291-5}

\bibitem{Krapivsky:2000}
P.~L. Krapivsky, S.~Redner, F.~Leyvraz, Connectivity of {{Growing Random
  Networks}}, Physical Review Letters 85~(21) (2000).

\bibitem{Dorogovtsev:2000}
S.~N. Dorogovtsev, J.~F.~F. Mendes, A.~N. Samukhin, Structure of {{Growing
  Networks}} with {{Preferential Linking}}, Physical Review Letters 85~(21)
  (2000).

\bibitem{Yule:1925}
F.~Y. E., G.~U. Yule, A {{Mathematical Theory}} of {{Evolution Based}} on the
  {{Conclusions}} of {{Dr}}. {{J}}. {{C}}. {{Willis}}, {{F}}.{{R}}.{{S}}.,
  Journal of the Royal Statistical Society 88~(3) (1925) 433.
\newblock \href {http://arxiv.org/abs/10.2307/2341419}
  {\path{arXiv:10.2307/2341419}}, \href {https://doi.org/10.2307/2341419}
  {\path{doi:10.2307/2341419}}.

\bibitem{Simon:1955}
H.~A. Simon, {{On a class of skew distribution functions}}, Biometrika 42~(3-4)
  (1955) 425--440.
\newblock \href {https://doi.org/10.1093/biomet/42.3-4.425}
  {\path{doi:10.1093/biomet/42.3-4.425}}.

\bibitem{Price:1976}
D.~D.~S. Price, \href{http://dx.doi.org/10.1002/asi.4630270505}{A general
  theory of bibliometric and other cumulative advantage processes}, Journal of
  the American Society for Information Science 27~(5) (1976) 292–306.
\newblock \href {https://doi.org/10.1002/asi.4630270505}
  {\path{doi:10.1002/asi.4630270505}}.
\newline\urlprefix\url{http://dx.doi.org/10.1002/asi.4630270505}

\bibitem{Merton:1968}
R.~K. Merton, \href{http://dx.doi.org/10.1126/science.159.3810.56}{The
  {Matthew} effect in {Science}: The reward and communication systems of
  science are considered.}, Science 159~(3810) (1968) 56–63.
\newblock \href {https://doi.org/10.1126/science.159.3810.56}
  {\path{doi:10.1126/science.159.3810.56}}.
\newline\urlprefix\url{http://dx.doi.org/10.1126/science.159.3810.56}

\bibitem{Hayes:2002}
B.~T. Hayes, \href{https://api.semanticscholar.org/CorpusID:123614498}{Follow
  the {Money}}, American Scientist (2002).
\newline\urlprefix\url{https://api.semanticscholar.org/CorpusID:123614498}

\bibitem{Boghosian:2015}
B.~M. Boghosian, M.~Johnson, J.~Marcq,
  \href{https://api.semanticscholar.org/CorpusID:254697740}{An {H} theorem for
  {Boltzmann}’s equation for the {Yard-Sale Model} of asset exchange},
  Journal of Statistical Physics 161 (2015) 1339 -- 1350.
\newline\urlprefix\url{https://api.semanticscholar.org/CorpusID:254697740}

\bibitem{Plaszczynski:2022}
S.~Plaszczynski, G.~Nakamura, C.~Deroulers, B.~Grammaticos, M.~Badoual,
  \href{https://link.aps.org/doi/10.1103/PhysRevE.105.054151}{Levy geometric
  graphs}, Phys. Rev. E 105 (2022) 054151.
\newblock \href {https://doi.org/10.1103/PhysRevE.105.054151}
  {\path{doi:10.1103/PhysRevE.105.054151}}.
\newline\urlprefix\url{https://link.aps.org/doi/10.1103/PhysRevE.105.054151}

\bibitem{Krapivsky:2001}
P.~L. Krapivsky, S.~Redner, Organization of {Growing Random Networks}, Physical
  Review E 63~(6) (2001) 066123.
\newblock \href {https://doi.org/10.1103/PhysRevE.63.066123}
  {\path{doi:10.1103/PhysRevE.63.066123}}.

\bibitem{Mandelbrot:1975}
B.~Mandelbrot,
  \href{https://users.math.yale.edu/mandelbrot/web_pdfs/comptes_rendus_79.pdf}{"sur
  un modèle décomposable d'univers hiérarchisé: déduction des
  corrélations galactiques sur la sphère céleste."}, Comptes Rendus (Paris)
  280A (1975) 1551--1554.
\newline\urlprefix\url{https://users.math.yale.edu/mandelbrot/web_pdfs/comptes_rendus_79.pdf}

\bibitem{Mandelbrot:1983}
B.~{Mandelbrot}, {The Fractal Geometry of Nature}, Freeman, San Francisco,
  1983.

\bibitem{Mandelbrot:1960}
B.~Mandelbrot, The {{Pareto-Levy Law}} and the {{Distribution}} of {{Income}},
  International Economic Review 1~(2) (1960) 79.
\newblock \href {http://arxiv.org/abs/2525289} {\path{arXiv:2525289}}, \href
  {https://doi.org/10.2307/2525289} {\path{doi:10.2307/2525289}}.

\bibitem{Levy:1925}
P.~L{\'e}vy, {Calcul des probabilit{\'e}s}, {Gauthier-Villars}, {Paris}, 1925.

\bibitem{Levy:1954}
P.~L{\'e}vy, {Th{\'e}orie de l'addition des variables al{\'e}atoires},
  {Gauthier-Villars}, {Paris}, 1954.

\bibitem{Paul:2013}
W.~Paul, J.~Baschnagel, Stochastic {{Processes}}, {Springer International
  Publishing}, {Heidelberg}, 2013.
\newblock \href {https://doi.org/10.1007/978-3-319-00327-6}
  {\path{doi:10.1007/978-3-319-00327-6}}.

\bibitem{Khintchine:1938}
A.~Y. Khintchine, Limit Distributions for the Sum of Independent Random
  Variables, {O.N.T.I}, {Moscow (in russian)}, 1938.

\bibitem{Gnedenko:1954}
B.~V. Gnedenko, A.~N. Kolmogorov, Limit Distributions for Sums of Independent
  Random Variables, {Addison-Wesley}, {Reading, Mass.}, 1954.

\bibitem{Seshadri:1982}
V.~Seshadri, B.~J. West, Fractal {Dimensionality} of {{L{\'e}vy}} processes,
  Proceedings of the National Academy of Sciences 79~(14) (1982) 4501--4505.
\newblock \href {https://doi.org/10.1073/pnas.79.14.4501}
  {\path{doi:10.1073/pnas.79.14.4501}}.

\bibitem{Chechkin:2008}
A.~V. Chechkin, R.~Metzler, J.~Klafter, V.~Y. Gonchar, Introduction to the
  {{Theory}} of {{L{\'e}vy Flights}}, in: R.~Klages, G.~Radons, I.~M. Sokolov
  (Eds.), Anomalous {{Transport}}, {Wiley-VCH Verlag GmbH \& Co. KGaA},
  {Weinheim, Germany}, 2008, pp. 129--162.
\newblock \href {https://doi.org/10.1002/9783527622979.ch5}
  {\path{doi:10.1002/9783527622979.ch5}}.

\bibitem{Hughes:1981}
B.~D. Hughes, M.~F. Shlesinger, E.~W. Montroll, Random walks with self-similar
  clusters, Proceedings of the National Academy of Sciences 78~(6) (1981)
  3287--3291.
\newblock \href {https://doi.org/10.1073/pnas.78.6.3287}
  {\path{doi:10.1073/pnas.78.6.3287}}.

\bibitem{Ozella:2021}
L.~Ozella, D.~Paolotti, G.~Lichand, J.~P. Rodr{\'i}guez, S.~Haenni, J.~Phuka,
  O.~B. {Leal-Neto}, C.~Cattuto, Using wearable proximity sensors to
  characterize social contact patterns in a village of rural {{Malawi}}, EPJ
  Data Science 10~(1) (2021) 46.
\newblock \href {https://doi.org/10.1140/epjds/s13688-021-00302-w}
  {\path{doi:10.1140/epjds/s13688-021-00302-w}}.

\bibitem{Newman:2005}
M.~E.~J. Newman, Power laws, {{Pareto}} distributions and {{Zipf}}'s law,
  Contemporary Physics 46~(5) (2005) 323--351.
\newblock \href {http://arxiv.org/abs/cond-mat/0412004}
  {\path{arXiv:cond-mat/0412004}}, \href
  {https://doi.org/10.1080/00107510500052444}
  {\path{doi:10.1080/00107510500052444}}.

\bibitem{Jaynes:1957}
E.~T. Jaynes, Information {{Theory}} and {{Statistical Mechanics}}, Physical
  Review 106~(4) (1957) 620--630.
\newblock \href {https://doi.org/10.1103/PhysRev.106.620}
  {\path{doi:10.1103/PhysRev.106.620}}.

\bibitem{Montroll:1983}
E.~W. Montroll, M.~F. Shlesinger, Maximum entropy formalism, fractals, scaling
  phenomena, and 1/f noise: {{A}} tale of tails, Journal of Statistical Physics
  32~(2) (1983) 209--230.
\newblock \href {https://doi.org/10.1007/BF01012708}
  {\path{doi:10.1007/BF01012708}}.

\bibitem{Metzler:2000}
R.~Metzler, J.~Klafter, The random walk's guide to anomalous diffusion: A
  fractional dynamics approach, Physics Reports 339~(1) (2000) 1--77.
\newblock \href {https://doi.org/10.1016/S0370-1573(00)00070-3}
  {\path{doi:10.1016/S0370-1573(00)00070-3}}.

\bibitem{Song:2010}
C.~Song, T.~Koren, P.~Wang, A.-L. Barab{\'a}si, Modelling the scaling
  properties of human mobility, Nature Physics 6~(10) (2010) 818--823.
\newblock \href {https://doi.org/10.1038/nphys1760}
  {\path{doi:10.1038/nphys1760}}.

\bibitem{Rybski:2009}
D.~Rybski, S.~V. Buldyrev, S.~Havlin, F.~Liljeros, H.~A. Makse, Scaling laws of
  human interaction activity, Proceedings of the National Academy of Sciences
  106~(31) (2009) 12640--12645.
\newblock \href {https://doi.org/10.1073/pnas.0902667106}
  {\path{doi:10.1073/pnas.0902667106}}.

\bibitem{Zipf:1949}
G.~K. Zipf, Human Behavior and the Principle of Least Effort: An Introduction
  to Human Ecology, {Addison-Wesly Press, Inc.}, {Cambridge 42. Massachusetts},
  1949.

\bibitem{Delanoe:2015}
A.~Delano\"e, B.~Conein, \href{https://doi.org/10.4000/sociologies.5046}{Le
  contr{\^{o}}le de la forme des r{\'{e}}seaux par leurs membres~: le fil de
  discussion comme r{\'{e}}seau d'interaction}, {SociologieS} (May 2015).
\newblock \href {https://doi.org/10.4000/sociologies.5046}
  {\path{doi:10.4000/sociologies.5046}}.
\newline\urlprefix\url{https://doi.org/10.4000/sociologies.5046}

\bibitem{Heesen:2022}
R.~Heesen, M.~Fr{\"o}hlich, Revisiting the human `interaction engine':
  Comparative approaches to social action coordination, Philosophical
  Transactions of the Royal Society B: Biological Sciences 377~(1859) (2022)
  20210092.
\newblock \href {https://doi.org/10.1098/rstb.2021.0092}
  {\path{doi:10.1098/rstb.2021.0092}}.

\bibitem{Peitgen:2004}
H.-O. Peitgen, H.~J\"{u}rgens, D.~Saupe,
  \href{http://dx.doi.org/10.1007/b97624}{Chaos and Fractals}, Springer New
  York, 2004.
\newblock \href {https://doi.org/10.1007/b97624} {\path{doi:10.1007/b97624}}.
\newline\urlprefix\url{http://dx.doi.org/10.1007/b97624}

\bibitem{Bouchaud:1990}
J.-P. Bouchaud, A.~Georges, Anomalous diffusion in disordered media:
  {{Statistical}} mechanisms, models and physical applications, Physics Reports
  195~(4-5) (1990) 127--293.
\newblock \href {https://doi.org/10.1016/0370-1573(90)90099-N}
  {\path{doi:10.1016/0370-1573(90)90099-N}}.

\bibitem{Jespersen:1999}
S.~Jespersen, R.~Metzler, H.~C. Fogedby, L{\'e}vy flights in external force
  fields: {{Langevin}} and fractional {{Fokker-Planck}} equations and their
  solutions, Physical Review E 59~(3) (1999) 2736--2745.
\newblock \href {https://doi.org/10.1103/PhysRevE.59.2736}
  {\path{doi:10.1103/PhysRevE.59.2736}}.

\bibitem{West:1994}
B.~J. West, W.~Deering, Fractal physiology for physicists: {{L{\'e}vy}}
  statistics, Physics Reports 246~(1-2) (1994) 1--100.
\newblock \href {https://doi.org/10.1016/0370-1573(94)00055-7}
  {\path{doi:10.1016/0370-1573(94)00055-7}}.

\bibitem{Graph-tool:2014}
T.~P. Peixoto, \href{http://figshare.com/articles/graph_tool/1164194}{The
  graph-tool python library}, figshare (2014).
\newblock \href {https://doi.org/10.6084/m9.figshare.1164194}
  {\path{doi:10.6084/m9.figshare.1164194}}.
\newline\urlprefix\url{http://figshare.com/articles/graph_tool/1164194}

\end{thebibliography}


\begin{thebibliography}{1}

\bibitem{Plaszczynski:2022}
S.~Plaszczynski, G.~Nakamura, C.~Deroulers, B.~Grammaticos, and M.~Badoual.
\newblock Levy geometric graphs.
\newblock {\em Phys. Rev. E}, 105:054151, May 2022.

\bibitem{Cattuto:2010}
Ciro Cattuto, Wouter {Van den Broeck}, Alain Barrat, Vittoria Colizza,
  Jean-Fran{\c c}ois Pinton, and Alessandro Vespignani.
\newblock Dynamics of {{Person-to-Person Interactions}} from {{Distributed RFID
  Sensor Networks}}.
\newblock {\em PLoS ONE}, 5(7):e11596, July 2010.

\bibitem{Barrat:2014}
A.~Barrat, C.~Cattuto, A.E. Tozzi, P.~Vanhems, and N.~Voirin.
\newblock Measuring contact patterns with wearable sensors: Methods, data
  characteristics and applications to data-driven simulations of infectious
  diseases.
\newblock {\em Clinical Microbiology and Infection}, 20(1):10--16, January
  2014.

\bibitem{Vanhems:2013}
Philippe Vanhems, Alain Barrat, Ciro Cattuto, Jean-Fran{\c c}ois Pinton, Nagham
  Khanafer, Corinne R{\'e}gis, Byeul-a Kim, Brigitte Comte, and Nicolas Voirin.
\newblock Estimating {{Potential Infection Transmission Routes}} in {{Hospital
  Wards Using Wearable Proximity Sensors}}.
\newblock {\em PLoS ONE}, 8(9):e73970, September 2013.

\bibitem{Isella:2010}
Lorenzo Isella, Juliette Stehl{\'e}, Alain Barrat, Ciro Cattuto, Jean-Fran{\c
  c}ois Pinton, and Wouter {Van den Broeck}.
\newblock What's in a crowd? {{Analysis}} of face-to-face behavioral networks.
\newblock {\em Journal of Theoretical Biology}, 271(1):166--180, December 2010.

\bibitem{Ozella:2021}
Laura Ozella, Daniela Paolotti, Guilherme Lichand, Jorge~P. Rodr{\'i}guez,
  Simon Haenni, John Phuka, Onicio~B. {Leal-Neto}, and Ciro Cattuto.
\newblock Using wearable proximity sensors to characterize social contact
  patterns in a village of rural {{Malawi}}.
\newblock {\em EPJ Data Science}, 10(1):46, December 2021.

\bibitem{Gelardi:2020}
Valeria Gelardi, Jeanne Godard, Dany Paleressompoulle, Nicolas Claidiere, and
  Alain Barrat.
\newblock Measuring social networks in primates: wearable sensors versus direct
  observations.
\newblock {\em Proceedings of the Royal Society A: Mathematical, Physical and
  Engineering Sciences}, 476(2236):20190737, 2020.

\end{thebibliography}

\end{document}


\bigskip
\title{ 
A stochastic model of discussion
\\\textbf{Supplementary Material}}
\author{ S. Plaszczynski, B. Grammaticos, M. Badoual}
\maketitle
\tableofcontents
\newpage

\section{The \textit{sociopatterns} datasets}

As detailed in \cite{Plaszczynski:2022}, we focus on 4 datasets from the \textit{sociopatterns}
collaboration that we consider as most dissimilar in terms of
sociological environment. More complete descriptions are available in the cited references.

The \textit{socopatterns} collaboration measure interactions between
(identified) individuals using RFID proximity sensors \cite{Cattuto:2010,Barrat:2014}.
Data is recorded for each pair of participants if they engage in a face-to-face exchange for
more than 20 s, which is also the resolution step of the instrument.
Data are publicly available from \safeurl{www.sociopatterns.org}

\begin{enumerate}
\item \hosp : data collected for 3 days \footnote{we will only
    consider complete (24 h) day periods.} on 75 
  participants in the geriatric unit of a
  hospital in Lyon (France) \cite{Vanhems:2013}. Most
  interactions (75\%) involve nurses and patients.
\item \conf: data taken at the ACM Hypertext 2009 (\safeurl{www.ht2009.org})
  conference with around a hundred participants (3 days)
  \cite{Isella:2010}
\item \malawi: data were taken in a small village Malawi (Africa) 
  with 86 participants for
  13 days. The group consists essentially of farmers \cite{Ozella:2021}.
\item \baboons: data taken at a Primate Center near Marseille
  (France) with 13 baboons for 26 days \cite{Gelardi:2020}.
\end{enumerate}

\section{Number of clusters in Levy graphs}
\label{sec:A}

As shown in \cite{Plaszczynski:2022}, the mean fraction of clusters
follows a power-law function of the scale with a relatively small
spread among realizations
\begin{align}
\label{eq:Nclus}
  \dfrac{\Nclus}{N}=\dfrac{A_c}{s^{\alpha_c}}
\end{align}

\begin{figure}[ht!]
  \centering
  \includegraphics[width=0.7\textwidth]{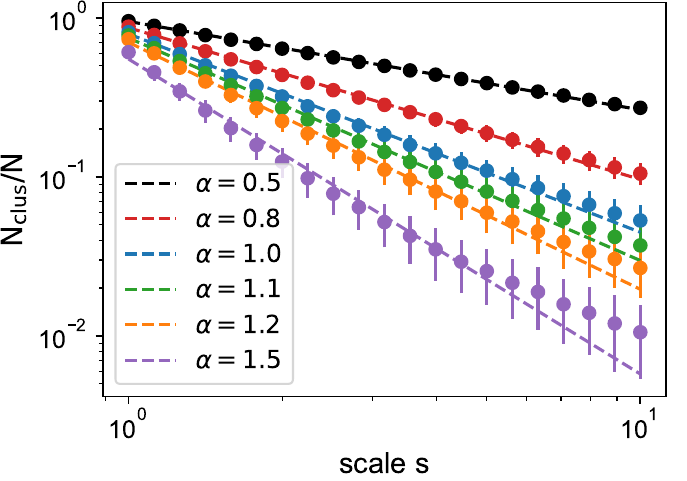}
\caption{\label{fig:fit}Mean fraction of clusters as a function of the
scale as determined from simulations for several $\alpha$ values.
The dashed lines represent the best power-law fits.}
\end{figure}

Although we give in the paper the exact formula for the coefficients
$A_c(\alpha)$ and $\alpha_c(\alpha)$ it still relies on several
coefficients (namely $c(\alpha),\beta(\alpha)$ and $\gamma(\alpha)$)
that must be determined from simulations.
Instead, we measure here directly their value from the simulations
using 1000 realizations of Levy graphs 
(N=100 000) with fixed $\alpha$ indices and varying the scale $s$ between 1 and 10.
\Fig{fit} shows the results and the power-law fits. The resulting
parameters are given in Table \ref{tab:fit}

\begin{table}
  \centering
\begin{tabular}{rcc}
\toprule
  $\alpha$ &    $A_c$ &   $\alpha_c$ \\
\midrule
0.50 & $0.95 \pm0.01$ & $0.56\pm0.01$ \\
0.80 & $0.87\pm0.02$ & $0.95\pm0.02$ \\
1.00 & $0.79\pm0.02$ & $1.25\pm0.04$ \\
1.10 & $0.75\pm0.02$ & $1.40\pm0.05$ \\
1.20 & $0.70\pm0.03$ & $1.55\pm0.06$ \\
1.50 & $0.55\pm0.03$ & $1.98\pm0.09$ \\
\bottomrule
\end{tabular}
\caption{\label{tab:fit} Coefficients of \refeq{Nclus} fitted 
  from the 2D Levy graph simulations (\Fig{fit}) for various $\alpha$ values.}
\end{table}

\section{Amplitude term coefficients}

The amplitude term $A$ in the distribution of the cluster size is
parametrized by
\begin{align}
\label{eq:beta}
  A(s)=c/s^\beta.
\end{align}

\begin{figure}[ht!]
  \centering
  \includegraphics[width=.5\linewidth]{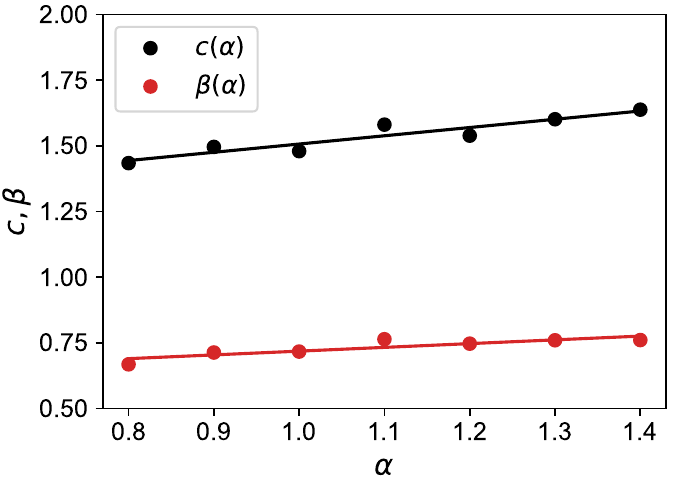}
  \caption{Determination of the values of the $c$ and $\beta$ parameters in the
    $A(s)=c/s^\beta$ parametrization 
    measured from simulations. The lines show the best linear fits.}
  \label{fig:ab_a}
\end{figure}

We use simulations to determine $c$ and $\beta$ and show the results in
dimension 2 on \Fig{ab_a}.

\section{Levy clusters distribution in any dimension}

Levy flights may evolve in a space of any dimension $d$. One just
needs to adapt the angular isotropic part of the jump to the space (a simple
algorithm to do it is given in \cite{Plaszczynski:2022}).
Although performed in a $d=2$ space, most computations in the paper
may be adapted to any dimension.

The cluster size distribution is still of the form
\begin{align}
\label{eq:pn}
 p(n)\propto {\prod_{k=1}^{n}(1-\dfrac{A}{k^\gamma})}
\end{align}

but now, as explained in the paper, we expect
\begin{align}
  \gamma\simeq\alpha/d
\end{align}

\begin{figure}[ht!]
  \centering
  \includegraphics[width=0.5\textwidth]{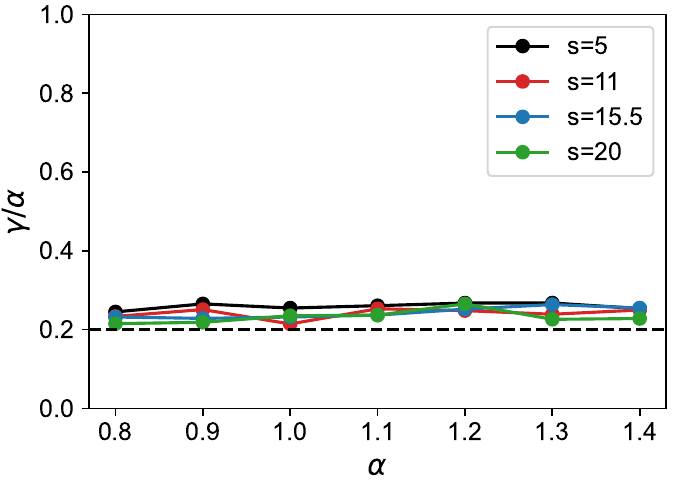}
\caption{\label{fig:gam_dim5} Estimate of the $\gamma$ parameter from
  the least square fit of \refeq{pn} model to Levy graph simulations
  in a space of dimension $d=5$. The dashed line corresponds to the
  naive $1/d$ expectation.}
\end{figure}

This is indeed what we observe at large scales in $d=5$ simulations as
shown in \Fig{gam_dim5}.

\begin{figure}[ht!]
  \centering
  \includegraphics[width=0.5\textwidth]{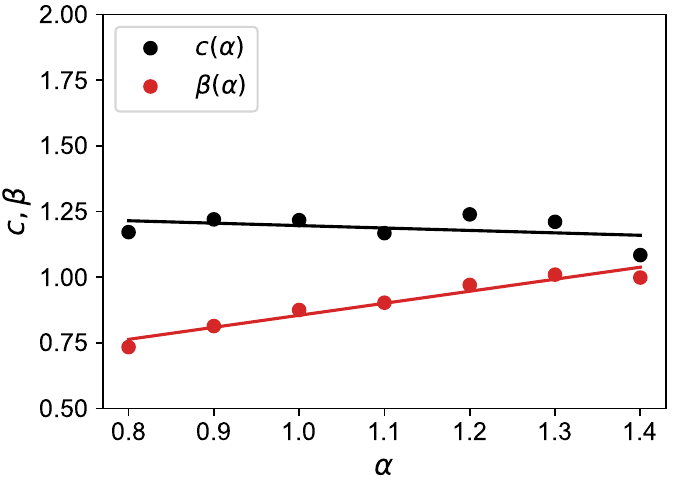}
\caption{\label{fig:a_cb_dim5} Estimate of the $c$ and $\beta$ parameters from
  the least square fit of \refeq{pn} model to Levy graph simulations
  in a space of dimension $d=5$.}
\end{figure}

As in dimension 2, the amplitude term $A$ is related to the scale
through
\begin{align}
  A=c/s^\beta,
\end{align}
but this time the $c$ and $\beta$ parameters dependency on $\alpha$
changes with the dimension.
For $d=5$ we find from simulations (\Fig{a_cb_dim5}) )
\begin{align}
  c&=1.28-0.09\alpha \\
  \beta&=0.40+0.45\alpha.
\end{align}

If we compare those results to the $d=2$ case (\Fig{ab_a}) we see that
the $c$ dependency on $\alpha$ has almost vanished while the $\beta$ one increased.
Asymptotically  we expect $\gamma\to0,~A\to1$ and $\beta\to \alpha$ in
order to match the geometric distribution. 

The asymptotic approximation for the probability of the cluster size is still valid
\begin{align}
\label{eq:asym}
  p(n) \propto \exp{\left(-\tfrac{A}{1-\gamma}n^{1-\gamma}\right)}, 
\end{align}
and the CCDF of the cluster size contrast is still 
\begin{align}
\label{eq:ccdf}
  p(\delta>x)=\dfrac{\Gamma[1/\mu,\left(x \Gamma(2/\mu)/\Gamma(1/\mu)\right)^\mu]}{\Gamma(1/\mu)}
\end{align}
but using this time
\begin{align}
  \mu=1-\alpha/d.
\end{align}

\begin{figure}[ht!]
  \centering
  \subfigure[]{\includegraphics[width=0.7\textwidth]{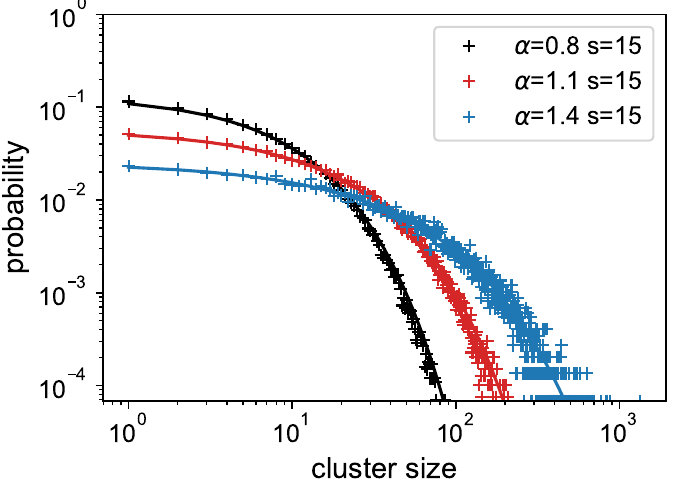}}
  \subfigure[]{\includegraphics[width=0.7\textwidth]{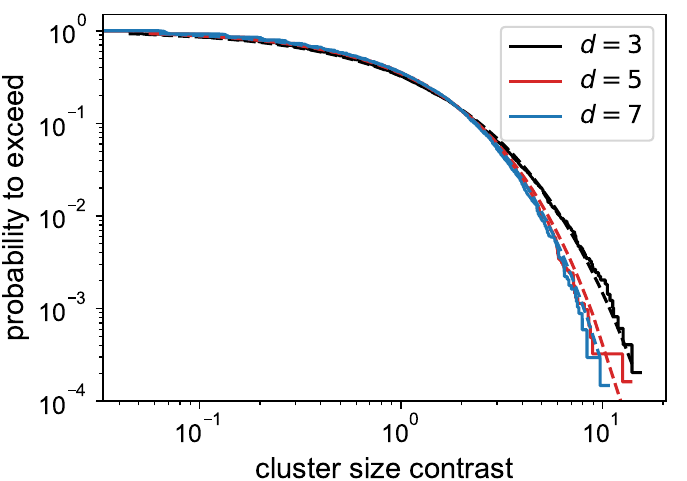}}
\caption{\label{fig:dims} (a) Distribution of the Levy clusters size in a
  space of dimension $d=5$. Points are obtained from simulations and
  the line corresponds to the analytical formula \refeq{asym}. CCDF of
the Levy cluster size contrast varying the dimension (for $\alpha=1.1$). The full lines are
obtained from simulation while the dashed ones represent the \refeq{ccdf} formula.}
\end{figure}

We verify both distributions using simulations on \Fig{dims}.

\section{The Maximum Likelihood Estimator}

Thre refined distribution of the cluster size contrast of Levy graphs with
index $\alpha$ above some cutoff $\delta_{min}$ is 
\begin{align}
  p(\delta;\alpha)&=\dfrac{g \mu}{\Gamma(1/\mu,\left(g
                    \delta_{min}\right)^{1/\mu})} e^{(-g\delta)^\mu}\\
  \text{with}~\mu=&1-\alpha/2,\quad g=f \frac{\Gamma(2/\mu)}{\Gamma(1/\mu)}.
  \label{eq:f}
\end{align}

The cutoff is important to be sensitive to the tail of the
distribution. Indeed the likelihood
treats ``equally'' all the data points and without it the Levy index
is mostly determined by low value samples which are the most numerous
in such a distribution.

For a given dataset with samples $\delta_i>\delta_{min}$, the
likelihood is
\begin{align}
\label{eq:lkl}
  \mathscr{L}(\alpha)=\prod_{i} p(\delta_i,\alpha).
\end{align}
  
By maximing it w.r.t $\alpha$ , which is traditionally performed by
minimizing $-\log \mathscr{L}$, one obtains the Maximum Likelihood
Estimate (MLE) $\hat \alpha$.

The factor $f$ that appears in Eq.(\ref{eq:f}) is determined from
simulations in such a way that the MLE is unbiased.
Each dataset defines a set of Levy graphs (of varying size and scale).
We fix the index of those graphs to some known value $\alpha_{true}$ and run 100
realizations per dataset. Each one gives a value of the MLE and we fix
$f$ in order that the mean of the observed values coincide with the
input $\alpha$ value. There is a slight dependency on $\alpha_{true}$
and by interpolation we obtain a function $f(\alpha)$ that ensures
that the MLE is always unbiased. 
 
\begin{figure}[ht!]
  \centering
  \subfigure[\malawi]{\includegraphics[width=0.49\textwidth]{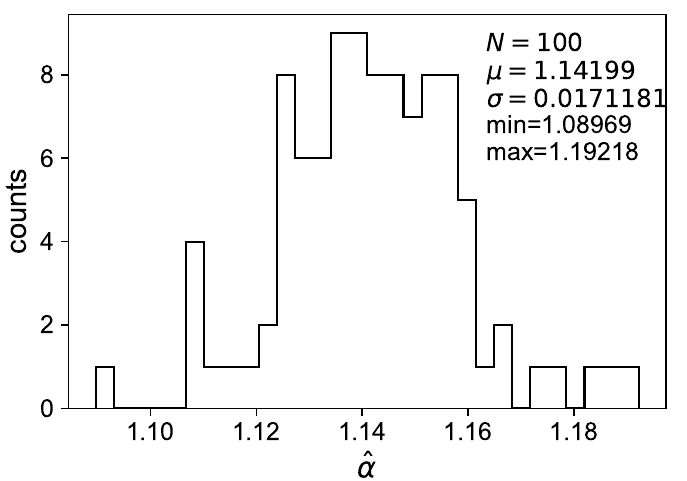}}
  \subfigure[\hosp]{\includegraphics[width=0.49\textwidth]{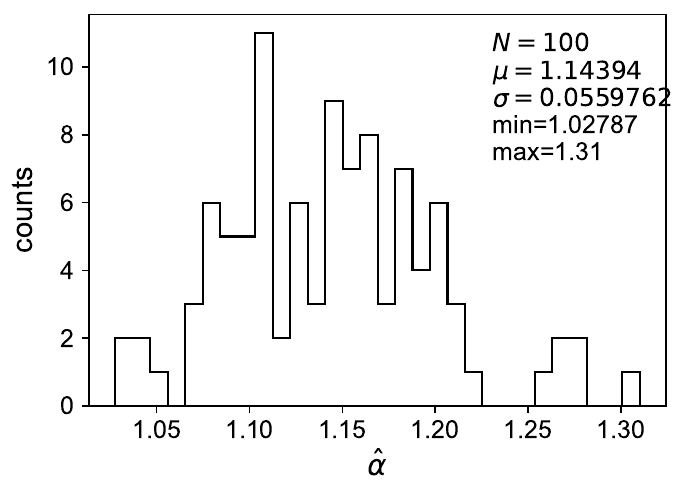}}\\
  \subfigure[\conf]{\includegraphics[width=0.49\textwidth]{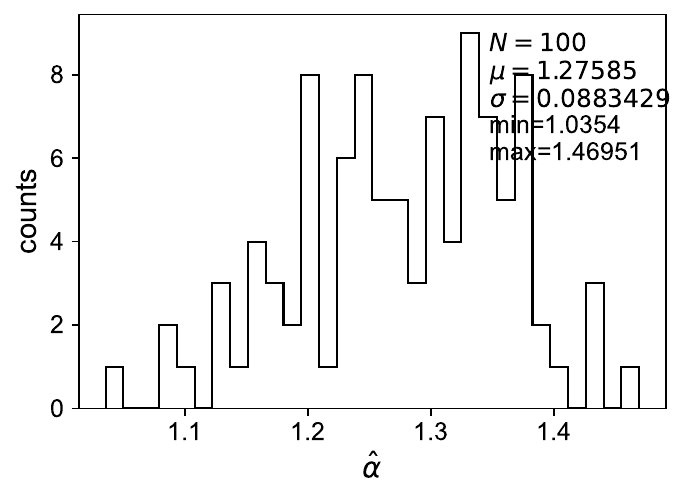}}
  \subfigure[\baboons]{\includegraphics[width=0.49\textwidth]{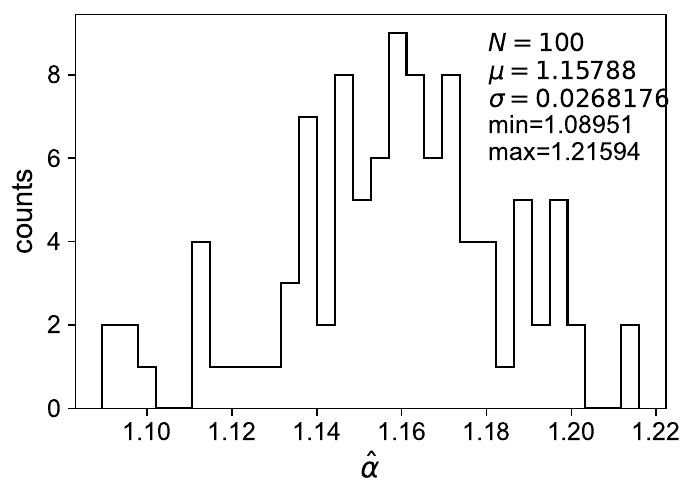}}\\
\caption{\label{fig:mle} Distributions of the Levy index MLE from
  simulations. The structure of the graphs
  is taken from each dataset and the Levy index fixed to
  $\alpha_{true}=1.14 (a) ,1.14(b) ,1.27(c), 1.16(d)$. The
  nomalization factor $f$ that appears in \refeq{f} ensures that the
  mean of the distributions coincides with $\alpha_{true}$, i.e the
  MLE is unbiased. Those distributions allow to obtain the precision of
  the estimator.}
\end{figure}

We show ain \Fig{mle} an example of the MLE distribution for each dataset with
$\alpha_{true}$ corresponding to the measured data values. The factor
$f$ ensures the mean of those distributions is correct.
From the distributions one obtains the standard deviation of the
estimator and from the 5 and 95\% percentiles a 95\% confidence level interval.

\section{Why dimension 2?}

\begin{figure}[ht!]
  \centering
  \includegraphics[width=0.7\textwidth]{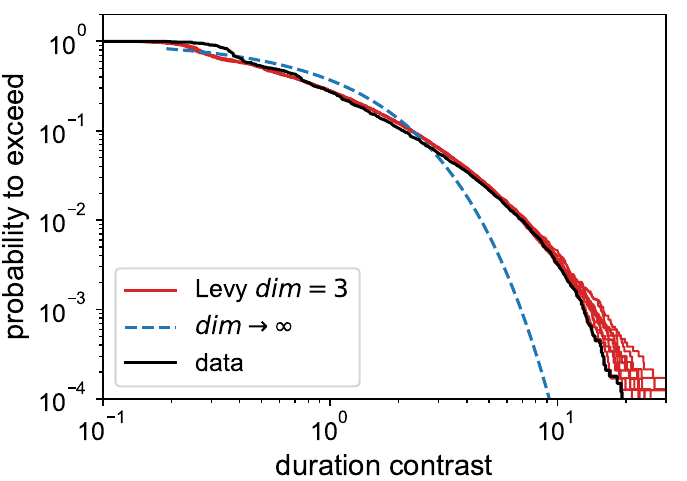}
\caption{\label{fig:dim3} Levy model ($\alpha=1.1$) in red compared to
the \malawi data (in black) for a Levy walk in a space of dimension 3.
Increasing the dimension further, the model converges to the dashed
blue line. }
\end{figure}

As stated in the paper, the choice of dimension 2 for the Levy walk is
only dictated by the data.
We have tried other dimensions. \Fig{dim3} shows the result in dimension
3 (using updated $A_c$ and $\alpha_c$ coefficients from \sect{A}). 
Compared to dimension 2 (see paper), the
agreement for the \malawi dataset gets worse in particular at low contrast. Increasing 
the dimension further degrades the agreement. Indeed the distribution
then moves to the geometric one, which has asymptotically
a $\exp(-\delta)$ form (dashed blue line in \Fig{dim3}) and is clearly far from the data.

\section{Levy flight self-similarity}

\begin{figure}[ht!]
  \centering
  \includegraphics[width=0.7\textwidth]{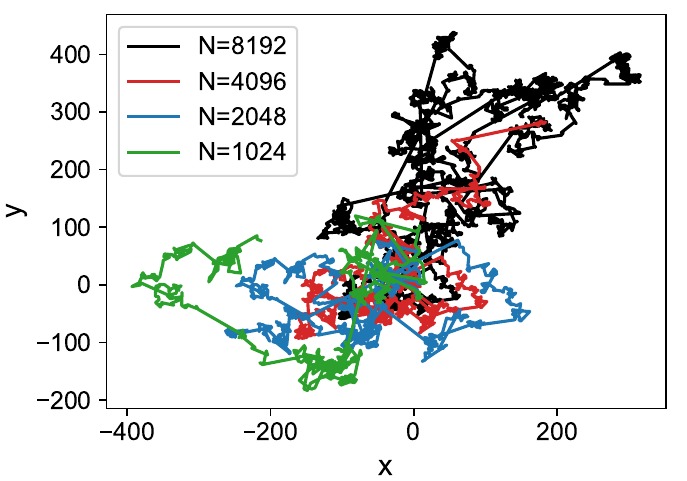}
\caption{\label{fig:self} Example of a Levy flight of index
  $\alpha=1.5$ with $N=8192$ points (in black). Then one takes the
  first half of the data ($N=4096$ points, in red) and rescale the coordinates
  by $2^H$ where $H=1/\alpha$. The procedure is repeated twice.}
\end{figure}

To illustrate what the Levy flight self-similarity means, we use
simulations.
We first generate a 2D Levy flight of index $\alpha=1.5$ with $N=8192$
points.
Then we take only the first half of the points but rescale the
coordinates by the factor 
\begin{align}
\label{eq:rescale}
  r=2^H
\end{align}
where the Hurst exponent is given by
\begin{align}
  H=1/\alpha
\end{align}
We repeat the procedure twice and show an example on \Fig{self}.

The trajectories of this statistical fractal ``look the same''. 
More precisely the $N$ points are drawn from the very same underlying distribution.

\newpage
\bibliographystyle{unsrt}
\bibliography{references}